\documentclass[12pt,draftcls, onecolumn]{IEEEtran}

\usepackage{amssymb}
\usepackage{amsmath}
\usepackage{amsfonts}
\usepackage{graphicx}
\usepackage{algorithm}
\usepackage{algorithmic}
\usepackage{epstopdf}
\usepackage{cite}
\usepackage{stfloats}
\usepackage{url}
\usepackage{CJK}
\usepackage{hyperref}
  \usepackage{epstopdf}
  \usepackage{subfigure}

  \usepackage{lineno}

\usepackage{algorithm,algorithmic}
\usepackage{multicol}

\usepackage{color,xcolor}

\newtheorem{theorem}{\text{Theorem}}

\newtheorem{example}{Example}

\newtheorem{lemma}{\text{Lemma}}

\renewcommand\thesubsection{\Alph{subsection}} 
\newtheorem{Prob}{Problem}

\begin{document}
\title{ {Joint Optimization of Preamble Selection and Access Barring for Random Access in MTC with General Device Activities}}

\author{\IEEEauthorblockN{Wang Liu, Ying Cui, Feng Yang, Lianghui Ding, and Jun Sun}\thanks{The authors are with Shanghai Jiao Tong University, China.
{This paper was presented in part at IEEE ICC Workshops 2021~\cite{LWICCWS21}.}}}



\maketitle

\begin{abstract}
Most existing random access schemes for machine-type communications (MTC) simply adopt a uniform preamble selection distribution, irrespective of the underlying device activity distributions. Hence, they may yield unsatisfactory access efficiency. In this paper, we model device activities for MTC as multiple Bernoulli random variables following an arbitrary multivariate Bernoulli distribution which can reflect both dependent and independent device activities. Then, we optimize preamble selection and access barring for random access in MTC according to the underlying joint device activity distribution. Specifically, we investigate three cases of the joint device activity distribution, i.e., the cases of perfect, imperfect, and unknown joint device activity distributions, and formulate the average, worst-case average, and sample average throughput maximization problems, respectively. The problems in the three cases are challenging nonconvex problems. In the case of perfect joint device activity distribution, we develop an iterative algorithm and a low-complexity iterative algorithm to obtain stationary points of the original problem and an approximate problem, respectively. In the case of imperfect joint device activity distribution, we develop an iterative algorithm and a low-complexity iterative algorithm to obtain a Karush-Kuhn-Tucker (KKT) point of an equivalent problem and a stationary point of an approximate problem, respectively. 
Finally, in the case of unknown joint device activity distribution, we develop an iterative algorithm to obtain a stationary point. The proposed solutions are widely applicable and outperform existing solutions for dependent and independent device activities.

\end{abstract}

\begin{IEEEkeywords}
 Machine-type communications (MTC), random access procedure, preamble selection, access barring, optimization.
\end{IEEEkeywords}
\newpage

\setlength{\abovedisplayskip}{2pt}
\setlength{\belowdisplayskip}{2pt}

\section{Introduction}

The emerging Internet-of-Things (IoT), whose key idea is to connect everything and everyone by the Internet, has received increasing attention in recent years.
Machine-type communications (MTC) are expected to support IoT services and applications, such as home automation, smart grids, smart healthcare, and environment monitoring~\cite{Yaacoub17IWC,hasan2013random}.
 Long Term Evolution (LTE) cellular networks offer the most natural and appealing solution for MTC due to ubiquitous coverage.
Specifically, the Third-Generation Partnership Project (3GPP) has developed radio technologies, such as LTE-M~\cite{LTEM_GC2017} and Narrowband IoT (NB-IoT)~\cite{NBIOT_ICM2017}, to enhance existing LTE networks and to provide new solutions inherited from LTE, respectively, for better serving IoT use cases.
In addition, 3GPP proposes to the International Telecommunications Union (ITU) that LTE-M and NB-IoT should be integrated as part of the fifth generation (5G) specifications~\cite{5GNR}.


In LTE-M and NB-IoT, devices compete in a random access channel (RACH) to access the base station (BS) through the random access procedure~\cite{TR_3gppevolved}.
Specifically, each active device randomly selects a preamble from a pool of available preambles according to a preamble selection distribution, and transmits it during the RACH.
The BS acknowledges the successful reception of a preamble if such preamble is transmitted by only one device.
In\text{\cite{Wong16TVT, Wong15TWC,  Kellerer19TWL ,Dailin18TWC, Dailin_18TWC2, QL_IOTJ2019, Seo_TVT2017,Koucheryavy_ISIT2013,Lee2015TWC, Choi_CL2016 ,Choi_WCNC2020}}, the authors consider the random access procedure and study the effect of preamble selection under certain assumptions on the knowledge of device activities.
Specifically,\text{\cite{Wong16TVT, Wong15TWC,  Kellerer19TWL , QL_IOTJ2019,Lee2015TWC, Choi_CL2016 ,Choi_WCNC2020}} assume that the number of active devices is known;~\cite{Koucheryavy_ISIT2013 , Seo_TVT2017 } assume that the distribution of the number of active devices is known;~\cite{Dailin18TWC, Dailin_18TWC2} assume that the statistics of the data queue of each device are known.
In\text{\cite{Wong16TVT, Wong15TWC,  Kellerer19TWL ,Dailin18TWC, Dailin_18TWC2, QL_IOTJ2019, Seo_TVT2017,Koucheryavy_ISIT2013,Lee2015TWC, Choi_CL2016,Choi_WCNC2020 }}, preambles are selected according to a uniform distribution, the average throughput~\cite{Wong16TVT, Wong15TWC,  Kellerer19TWL ,Dailin18TWC, Dailin_18TWC2, QL_IOTJ2019, Seo_TVT2017,Koucheryavy_ISIT2013,Lee2015TWC, Choi_CL2016,Choi_WCNC2020}, average access efficiency~\cite{Lee2015TWC} and resource consumption~\cite{ Kellerer19TWL} are analyzed and the number of allocated preambles is optimized to maximize the average throughput~\cite{Kellerer19TWL,Choi_CL2016 } or access efficiency~\cite{Lee2015TWC}.
Notice that optimal solutions are obtained in~\cite{Kellerer19TWL ,Lee2015TWC, Choi_CL2016}.
 In~\cite{  Popovski_PIMRC2020,Saad_TCOM2020,Saad_CL2018}, the active devices are assumed to be known.
	 Under this assumption, each active device is allocated a preamble if the number of preambles is greater than or equal to the number of active devices, and the preambles are allocated to a subset of active devices otherwise.

When many active devices attempt to access a BS simultaneously, a preamble is very likely to be selected by more than one device, leading to a significant decrease in the probability of access success.
In this scenario, access control is necessary.
One widely used access control method is the access barring scheme~\cite{TR_RanImprovement}, which has been included in the LTE specification in~\cite{TR_3gppevolved}.
In\text{\cite{Wong16TVT, Wong15TWC,  Kellerer19TWL ,Dailin18TWC, Dailin_18TWC2, QL_IOTJ2019, Seo_TVT2017,Koucheryavy_ISIT2013,Lee2015TWC}}, the authors consider access barring, besides random access procedure with preambles selected according to a uniform distribution.
Specifically, the access barring factor is optimized to maximize the average throughput~\cite{ Wong16TVT, Wong15TWC,  Kellerer19TWL ,Dailin18TWC, Dailin_18TWC2, QL_IOTJ2019, Seo_TVT2017} or access efficiency~\cite{Lee2015TWC}, or to minimize the number of backlogged devices~\cite{Koucheryavy_ISIT2013}.
Note that optimal solutions are obtained in~\cite{Koucheryavy_ISIT2013, Wong16TVT, QL_IOTJ2019,Kellerer19TWL , Seo_TVT2017,  Lee2015TWC }, an approximate solution is obtained in~\cite{Wong15TWC}, and asymptotically optimal solutions for a large number of devices are obtained in~\cite{Dailin18TWC,Dailin_18TWC2}.

	There exist two main types of MTC traffic, i.e., event-driven (or event-triggered) traffic and periodic traffic~\cite{ET-R1,ET-R2}.
In event-driven MTC (e.g., smart health care and environment sensing such as rockfall detection and fire alarm), one device being triggered may increase the
probability that other devices in the vicinity trigger in quick succession, or many devices are simultaneously triggered in response to a particular IoT event.
 Hence, device activities can be dependent.
Besides, in periodic MTC, devices have their own periods, and hence device activities can be considered independent for tractability (if the periodic behaviors of devices are not tracked).
Intuitively, effective preamble selection and access barring should adapt to the statistical behaviors of device activities.
Specifically, 
if devices are more likely to activate simultaneously, letting them always select different preambles can avoid more collisions; if devices are less likely to activate simultaneously, letting them always select the same preamble can avoid more collisions.
However, most existing random access schemes simply adopt a uniform preamble selection distribution, irrespective of the underlying device activity distributions~\cite{Wong16TVT, Wong15TWC,  Kellerer19TWL ,Dailin18TWC, Dailin_18TWC2, QL_IOTJ2019, Seo_TVT2017,Koucheryavy_ISIT2013,Lee2015TWC, Choi_CL2016,Choi_WCNC2020}.
		 They hence may yield far from optimal access efficiency.
This motivates us to
 model device activities as multiple Bernoulli random variables following an arbitrary multivariate Bernoulli distribution, which can reflect both dependent and independent device activities~\cite{cui2020jointly, jiang2021ml, Popovski18SPAWC, TSP_LL1_2018}, and optimize preamble selection and access barring according to the multivariate Bernoulli distribution~\cite{Popovski18SPAWC}.
To our knowledge, \cite{Popovski18SPAWC} is the first work that considers a general joint device activity distribution and
attempts to optimize the preamble selection distributions and access barring factors of all devices.
More specifically, in \cite{Popovski18SPAWC}, the authors assume that a perfect joint device activity distribution is known, maximize an approximation of the average throughput, which captures the active probabilities of every single device and every two devices, and develop a heuristic algorithm to tackle the challenging nonconvex problem.
The approximation error and the heuristic algorithm may yield a nonnegligible loss in access efficiency.
Therefore, it is critical to explore more effective algorithms adaptive to statistics of device activities.
Furthermore, notice that most existing works assume that the number of active devices~\cite{Wong16TVT,  Wong15TWC,Kellerer19TWL,  QL_IOTJ2019,Lee2015TWC,Choi_CL2016}, the distribution of the number of active devices~\cite{Seo_TVT2017, Koucheryavy_ISIT2013}, the statistics of the data queue of each device~\cite{Dailin18TWC, Dailin_18TWC2}, the active devices\text{~\cite{ Popovski_PIMRC2020,Saad_TCOM2020,Saad_CL2018}}, or the device activity distribution~\cite{Popovski18SPAWC} is perfectly known.
However, in practice, such assumptions are hardly satisfied.
Hence, it is critical to consider the optimization of preamble selection and access barring under imperfect information of device activities or even historical samples of device activities.

This paper investigates the joint optimization of preamble selection and access barring for IoT devices in MTC, whose activities are modeled by an arbitrary multivariate Bernoulli distribution, which can reflect both dependent and independent device activities.  
Specifically, we consider three cases of the joint device activity distribution, i.e., the case of perfect general joint device activity distribution (where the joint device activity distribution has been perfectly estimated), the case of imperfect joint device activity distribution (where the estimation error of the joint device activity distribution lies within a known deterministic bound), and the case of unknown joint device activity distribution (where activity samples generated according to the joint device activity distribution are available).
The optimization-based solutions are generally applicable for any dependent and independent device activities.
Our main contributions are summarized below.
\begin{itemize}
\item
In the case of perfect joint device activity distribution, we formulate the average throughput maximization problem, which is nonconvex with a complicated objective function. Based on the block coordinate descend (BCD) method~\cite{Bertsekas1998NP}, we develop an iterative algorithm, where most block coordinate optimization problems are solved analytically, and a low-complexity iterative algorithm, where all block coordinate optimization problems have closed-form optimal points, to obtain stationary points of the original problem and an approximate problem, respectively. Furthermore, we characterize an optimality property of a globally optimal point of the original problem.
\item
In the case of imperfect joint device activity distribution, we formulate the worst-case average throughput maximization problem as a robust optimization problem, which is a max-min problem. By duality theory and successive convex approximation (SCA)~\cite{razaviyayn2014successive}, we develop an iterative algorithm to obtain a Karush-Kuhn-Tucker (KKT) point of an equivalent problem of the max-min problem. Based on the BCD method, we also develop a low-complexity iterative algorithm, where all block coordinate optimization problems are solved analytically, to obtain a stationary point of an approximate problem of the max-min problem.
\item
In the case of unknown joint device activity distribution, we formulate the sample average throughput maximization problem, which can be viewed as a nonconvex stochastic optimization problem. Based on mini-batch stochastic parallel SCA~\cite{koppel2018parallel}, we develop an efficient parallel iterative algorithm where the approximate convex problems in each iteration are solved analytically, to obtain a stationary point.
\item
Finally, we show that for dependent and independent device activities, the proposed solutions achieve significant gains over existing schemes in all three cases by numerical results.
Besides, the performance of each low-complexity algorithm is close to that of its counterpart designed for solving the original problem.
\end{itemize}
The key notation used in this paper is listed in Table I.
 \begin{table}[t]
	\caption{Key Notation}
	\centering
	\begin{tabular}{|l|l|l|l|}\hline
		Notation&Description & 	Notation &Description \\ \hline
		$K$& the number of devices  &     	$a_{k,n}\in [0,1]$ &     the probability that device $k$ selects preamble $n$   \\ \hline
		$N$&  the number of preambles  &  	$ 	T({\bf A},{\epsilon},{\bf x})$  &average throughput conditional on $\mathbf x$    \\ \hline
		$x_k \in \{0,1\}$&  the activity state of device $k$      &  $ \bar	T({\bf A},{\epsilon},{\bf p})$  & average throughput  \\ \hline
		$ p_{\bf x} \in [0,1]$  & the probability that the activity states are $\bf x$   & 	$ \tilde	T ({\bf A},{\epsilon},{\bf p})$     & approximate average throughput   \\\hline
			$ \underline p_{\bf x}  \in [0,1]$  & the lower bound of $p_{\bf x}$  & $\bar T_{\text{wt}}({\bf A},{\epsilon},\mathcal P) $       &  worst-case average throughput  \\\hline
				$ \overline p_{\bf x}\in [0,1]$  & the upper bound of $p_{\bf x}$  &  $\tilde T_{\text{wt}}({\bf A},{\epsilon}, \tilde{ \mathcal P}) $    & approximate worst-case average throughput   \\\hline
		$\epsilon \in [0,1]$	& the access barring factor  &         $ \bar T_{\text{st}}({\bf A},{\epsilon})  $  & sample average throughput \\\hline	
	\end{tabular}
	\label{tab:Margin_settings}
\end{table}


\section{System Model }
We consider the uplink of a single-cell wireless network consisting of one BS and $K$ devices.
Let $\mathcal K \triangleq \{1,2,...,K\}$ denote the set of $K$ devices.
We consider a discrete-time system with time being slotted and assume that devices activate independently and identically over slots within a certain period.
In each slot, a device can be either active or inactive. Thus, the activities of the $K$ devices in a slot can be modeled as $K$ Bernoulli random variables, following an arbitrary multivariate Bernoulli distribution~\cite{TSP_LL1_2018,cui2020jointly,jiang2021ml}.
Generally, the $K$ Bernoulli random variables (representing the activities of the $K$ devices) can be dependent or independent.
Let $x_k \in \{0,1\}$ denote the activity state of device $k$, where $x_k=1$ if device $k$ is active and $x_k=0$ otherwise.
Let binary vector ${\bf x}\triangleq (x_{k})_{k\in \mathcal K}  \in  \mathcal X $ denote the activity states of the $K$ devices, where $  \mathcal X \triangleq \{0,1\}^K$.
The joint device activity distribution (assumed invariant over the considered period)
 is denoted by ${\bf p}\triangleq (p_{\bf x})_{{\bf x}\in \mathcal X} $, 
  where $p_{\bf x}$ represents the probability of the activity states of the $K$ devices being $\bf x$.
 Note that whether the device activities are dependent or independent depends on the values or forms of $\mathbf p$'s components.  
According to the nonnegativity and normalization axioms, we have
	\begin{subequations}\label{C_DAP}
		\begin{align}
			&    p_{\bf x} \geq 0, \  {\bf x} \in \mathcal X, \label{Cp1} \\
			& \sum\limits_{{\bf x} \in \mathcal X} p_{\bf x} =1.  \label{Cp2}
		\end{align}
	\end{subequations}
In this paper, we consider $ p_{\mathbf 0} < 1 $.\footnote{$ p_{\mathbf 0} =1 $ is not an interesting or practical scenario.}
We do not pose any additional assumption on $\mathbf p$.
	Thus, all analytical and optimization results, relying on $\mathbf p$ or its related quantities, are general.

In a slot, each active device tries to access the BS.
	Congestion may occur when many active devices
	require to access the BS at the same time.
	We adopt an access barring scheme for access control~\cite{TR_RanImprovement}.
	In particular, at the beginning of each slot, all active devices
	independently
attempt to access the BS with probability $\epsilon$, where
	\begin{subequations}\label{C_ACB}
		\begin{align}
			&\epsilon \geq 0, \label{C1} \\
			&\epsilon \leq1.  \label{C12}
		\end{align}
	\end{subequations}
	Here, $\epsilon$ is referred to as the access barring factor for all devices and will be optimized later.\footnote{The average throughput can be improved when devices have different access barring factors.
	To avoid resource starvation for each device and 	
	  maintain fairness, we consider identical access barring factors for all devices in the optimizations for simplicity.
	The proposed solution framework can be extended to handle different access barring factors.}
	That is, the access barring scheme is parameterized by the access barring factor $\epsilon$.


We adopt the random access procedure, which consists of four stages, i.e., \textit{preamble transmission}, \textit{random access response}, \textit{scheduled transmission}, and \textit{contention resolution}~\cite{hasan2013random}.
We focus only on the first stage, which mainly determines the success of access~\cite{Wong16TVT, Wong15TWC,  Kellerer19TWL ,Dailin18TWC, Dailin_18TWC2, QL_IOTJ2019, Seo_TVT2017,Koucheryavy_ISIT2013,Lee2015TWC,  Choi_CL2016,Choi_WCNC2020,  Popovski_PIMRC2020,Saad_TCOM2020,Saad_CL2018,Popovski18SPAWC}.
Consider $N$ orthogonal preambles, the set of which is denoted by $\mathcal N \triangleq \{1,2,...,N\}$.
Specifically, at the first stage, each device that attempts to access the BS independently selects a preamble out of the $N$ preambles to transmit.
The probability that device $k$ selects preamble $n$ is denoted by $a_{k,n}$,
which satisfies
\begin{subequations}\label{C_PSDs}
	\begin{align}
		& \ a_{k,n} \geq 0, \ k \in \mathcal K, n \in \mathcal N, \label{C2} \\
		&\sum\limits_{n \in \mathcal N} a_{k,n}= 1, \  k\in \mathcal K. \label{C3}
	\end{align}
\end{subequations}
Let ${\bf a}_{k} \triangleq  (a_{k,n})_{  n \in \mathcal N}$
denote the preamble selection distribution of device $k$.
Let ${\bf A}\triangleq ({ a}_{k,n})_{k \in \mathcal K, n\in \mathcal N}$ denote the distributions of the $K$ devices.
The $k$-th column of $\bf A$ is ${\bf a}_k$.
Note that for all $k \in \mathcal K$, the random \textit{preamble transmission} parameterized by ${\bf a} _k$ reduces to the \textit{preamble transmission} in the standard random access procedure \cite{TR_3gppevolved} when $a_{k,n} =\frac {1}{N}$, $ n \in \mathcal N$.
Furthermore, note that for all $k \in \mathcal K$, the considered preamble selection is generally random and becomes deterministic when $a_{k,n} \in \{0,1\}, n \in \mathcal N$.
We allow ${\bf a}_{k}$, $k \in \mathcal K$ to be arbitrary distributions to maximally avoid the collision.

If a preamble is selected by a single device, this device successfully accesses the BS \cite{Dailin18TWC}.
Then, the average number of devices that successfully access the BS at activity states $\bf x$ in a slot is given by~\cite{Popovski18SPAWC}
\begin{align}
	T({\bf A},{\epsilon},{\bf x})  \triangleq \sum\limits_{n\in \mathcal N}
	\sum\limits_{k\in \mathcal K} x_k a_{k,n} \epsilon \prod\limits_{\ell\in {\mathcal K:\ell \neq k}} (1-x_{\ell} a_{\ell,n}\epsilon ). \label{T_x}
\end{align}
In this paper, we consider the following three cases of joint device activity distribution and introduce the respective performance metrics.\newline
\textbf{Perfect joint device activity distribution}: In this case, we assume that the joint device activity distribution has been estimated by some learning methods, and the estimation error is negligible. That is, the exact value of $\bf p$ is known.
 We adopt the average throughput~\cite{Popovski18SPAWC}
   \begin{align}
 \bar T({\bf A},{\epsilon}, {\bf p} ) & \triangleq  \sum\limits_{{\bf x} \in \mathcal X}p_{\bf x}    T({\bf A},{\epsilon},{\bf x})   \label{T_def}  \\
&= \sum\limits_{m=1}^{K}
    m(-1)^{m-1} \epsilon^{m} \sum\limits_{n\in \mathcal N}\sum\limits_{
    \mathcal K' \subseteq {\mathcal K}:|\mathcal K' |=m}
   \left( \sum\limits_{{\bf x \in \mathcal X}}
p_{\bf x}
  \prod\limits_{k \in \mathcal K'}{x}_{k}\right) \prod\limits_{k \in \mathcal K'}  a_{k,n}    ,
   \label{T_Avg.New}
  \end{align}
as the performance metric, where $ T({\bf A},{\epsilon},{\bf x})$ is given by (\ref{T_x}). The proof for~(\ref{T_Avg.New}) can be found in Appendix A.
Note that for all $ \mathcal K' \subseteq \mathcal K $, $\sum\nolimits_{{\bf x \in \mathcal X}}$
$p_{\bf x}
  \prod\nolimits_{k \in \mathcal K'}{x}_{k}$ represents the probability that all devices in $ \mathcal K'$ are active and can be computed in advance. \newline
\textbf{Imperfect joint device activity distribution}: In this case, we assume that the joint device activity distribution has been estimated by some learning methods with certain estimation errors.
For all ${\bf x} \in \mathcal X$,
 {
 let $\hat p_{\bf x}$ denote the
 estimated probability of the device activity states being $\bf x$, and
 let $\Delta_{{\bf x}} \triangleq  p_{\bf x} -\hat p_{\bf x} $
 }
 denote the corresponding estimation error.
Note that
 $\Delta_{{\bf x}}, {\bf x} \in \mathcal X $ satisfy
 $\sum\nolimits_{{\bf x} \in   \mathcal X}  \Delta_{{\bf x} }  = 0 $, and $|\Delta_{{\bf x}} | \leq \delta_{\bf x}, {\bf x} \in \mathcal X$
for some known estimation error bounds $\delta_{{\bf x}} \in [0,1)$, ${\bf x} \in \mathcal X $.
Assume that $\hat p_{\bf x}$, $\delta_{{\bf x}}, {\bf x} \in \mathcal X $
 are known to the BS, but neither $\bf p$ nor $\Delta_{\bf x}, {\bf x} \in \mathcal X  $ is known to the BS.
  Thus, the BS knows that the exact joint
 activity distribution satisfies $\bf p \in \mathcal P$, where
 \begin{align}
 \mathcal P \triangleq
    \left\{ (y_{\bf x})_ {{\bf x}\in \mathcal X} \bigg|\   \underline{p}_{\bf x} \leq y_{\bf x} \leq \overline{p}_{\bf x} ,\ {\bf x} \in \mathcal X, \ \sum\limits_{{\bf x} \in \mathcal X}  y_{\bf x} =1 \right\},   \nonumber
 \end{align}
 with $\underline{p}_{\bf x} \triangleq \max\{\hat p_{\bf x} - \delta_{\bf x},0\}$ and
    $\overline{p}_{\bf x} \triangleq \min\{\hat p_{\bf x} + \delta_{\bf x} , \ 1\}$, for all ${\bf x}\in \mathcal X$. Note that $\mathcal P $ reflects $\mathbf p$ to a certain extent. It can be easily verified that $\mathcal P$ shrinks to $\{\mathbf p\}$ as the estimation error bounds $\delta_{\mathbf x}$, $\mathbf x \in \mathcal X$ decrease to $0$. 
We adopt the worst-case average throughput
 \begin{align}
  & \bar T_{\text{wt}}({\bf A},{\epsilon},\mathcal P) \triangleq \min_{{\bf p} \in \mathcal P} \bar T({\bf A},{\epsilon},{\bf p} ) \label{T_wt}
    \end{align}
as the performance metric, where $ \bar T({\bf A},{\epsilon},{\bf p} ) $ is given by (\ref{T_Avg.New}).
Note that $	\bar T_{\text{wt}}({\bf A},{\epsilon},\mathcal P ) \leq     \bar T({\bf A},{\epsilon}, {\bf p} )$ with the equality holds if and only if $\delta_{\mathbf x} = 0$, $\mathbf x \in \mathcal X$.\newline
\textbf{Unknown joint device activity distribution}:
In this case, we assume no direct information on the joint device
 activity distribution, but $I$ samples of the device activity states, generated according to $\mathbf p$, are available. 
 The $I$ samples reflect $\mathbf p$ to a certain extent via the empirical joint probability distribution, an unbiased estimator of $\mathbf p$. It can be easily verified that the mean squared error decreases with $I$.
 Denote $\mathcal I \triangleq \{1,2,...,I\} $.
For all $i \in \mathcal I$, the $i$-th sample is denoted by ${\bf x}_i$.
We adopt the sample average throughput
 \begin{align}
  & \bar T_{\text{st}}({\bf A},{\epsilon}) \triangleq
  \frac{1}{I} \sum\limits_{i \in \mathcal I}T \left({\bf A},{\epsilon},{\bf x}_i \right )    \label{T_st}
    \end{align}
as the performance metric, where $  T({\bf A},{\epsilon},{\bf x})$ is given by (\ref{T_x}).
Note that $	\bar T_{\text{st}}({\bf A},{\epsilon}  )  $ is a random variable 
	with randomness induced by the $I$ samples and $\mathbb E[	\bar T_{\text{st}}({\bf A},{\epsilon}  )] =\bar T ({\bf A},{\epsilon},\mathbf p ) $.
 
As illustrated in the introduction, overlooking the joint device activity distribution $\mathbf p$ (or related quantities, such as $\mathcal P$ and $\mathbf x_i$, $i\in\mathcal I$) may yield ineffective preamble selection and access barring designs.
	This can be further seen from the following example.

 \begin{example}[Motivation Example]
  	We consider $K = 3$ devices and $N=2$ preambles.
 The marginal probability of each device being active is $ p \in (0,\frac{1}{2}]$;
 the activities of device~1 and device~2 can be correlated (dependent),\footnote{Note that correlation (correlated) implies dependence (dependent).}
  and their joint distribution $p_{x_1,x_2}$, $x_1,x_2 \in \{0,1 \} $ is given by $ p_{0,0} = 1+ (\eta- 2)p + (1-\eta)p^2 $, $p_{0,1} =p_{1,0} =  (1-\eta)(p-p^2)$, and $p_{1,1 } =\eta p +  (1-\eta)p^2 $ \cite{jiang2021ml},
where $\eta \in [   \frac{p}{p -1}   , 1   ]$ represents the correlation coefficient;\footnote{The range of $\eta $ is to guarantee that $p_{x_1,x_2}$, $x_1,x_2 \in \{0,1 \} $ satisfy (1a)-(1b).} and the activity of device~3 is independent of the activities of device~1 and device~2.
 Thus, $\mathbf p $ is given by
$
 	p_{x_1,x_2,x_3} = p_{x_1,x_2}p_{x_3}, \  x_1,x_2,x_3 \in \{0,1\}
 $.
Note that $\eta \neq 0$ corresponds to dependent device activities, and $\eta = 0 $ corresponds to  independent device activities.
 We consider four feasible random access schemes parameterized by $(\mathbf A_i, \epsilon_i)$, $i = 1,2,3,4$.
 Specifically, $\mathbf A_1 = [\mathbf e_1,\mathbf e_1,\mathbf e_2]$, $\mathbf A_2 = [\mathbf e_1,\mathbf e_2,\mathbf e_2]$,
 \begin{align}
 	\mathbf A_3= \left\{  \begin{array}{ll}
 		\mathbf A_1 &, \  \  \eta \leq 0 \\
 		\mathbf A_2&, \    \  \eta >0  \end{array}
 	\right.  ,  \nonumber
\end{align}
and $\mathbf A_4 =\frac{1}{2} [\mathbf 1,\mathbf 1,\mathbf 1]$ (uniform distributions), where ${\bf e}_n \in \mathbb R^N$ represents the column vector of all zeros except the $n$-th entry being $1$, and $\mathbf 1  \in \mathbb R^N$ represents all-one column vector;
and
\begin{align}
	\epsilon_i = \mathrm{arg}\max_{\epsilon}  \bar T(\mathbf A_i, \epsilon,\mathbf p) =\left\{
	\begin{array}{ll}
			\min\left(1 ,  \frac{3}{ 4 } \frac{1}{\eta +  (1-\eta)p } \right)&, \     i =1 \\
			1 &, \   i =2,3,4
	\end{array}  \right.  \!\! .  \nonumber
\end{align} 
Note that $\mathbf A_1$, $\mathbf A_2$, and $\mathbf A_4$ do not rely on $\mathbf p$, whereas $\mathbf A_3$ depends on the parameter of $\mathbf p $, i.e., $\eta$.
\begin{figure}[t]
	\centering
	\includegraphics[width=0.4 \textwidth]{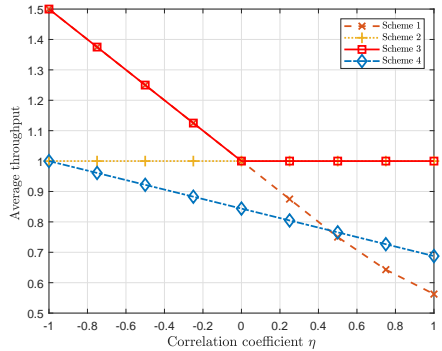}
	\vspace{-1mm}
	\caption{\small{Average throughput versus correlation coefficient $\eta$ for Example 1 at $p=0.5$.}}\label{Figure:Motivation}
	\vspace{-5mm}
\end{figure}
 Fig.~\ref{Figure:Motivation} plots the average throughput versus the correlation efficient $\eta$. 
 From Fig.~\ref{Figure:Motivation}, we can see that Scheme~3 outperforms Scheme~1 and Scheme~4 at $\eta>0$ and outperforms Scheme~2 and Scheme~4 at $\eta \leq 0$. This is because when device~1 and device~2 are more (or less) likely to activate simultaneously, letting them always select different preambles ({or the identical preamble}) can avoid more collisions.
 Besides, Scheme~3 outperforms Scheme~4 at $\eta = 0$.
 This example indicates that adapting $(\mathbf A, \epsilon)$ to $\mathbf p$ can improve the performance of the random access scheme.
\end{example}

\begin{figure}[t]
	\centering
	\includegraphics[width=1\textwidth]{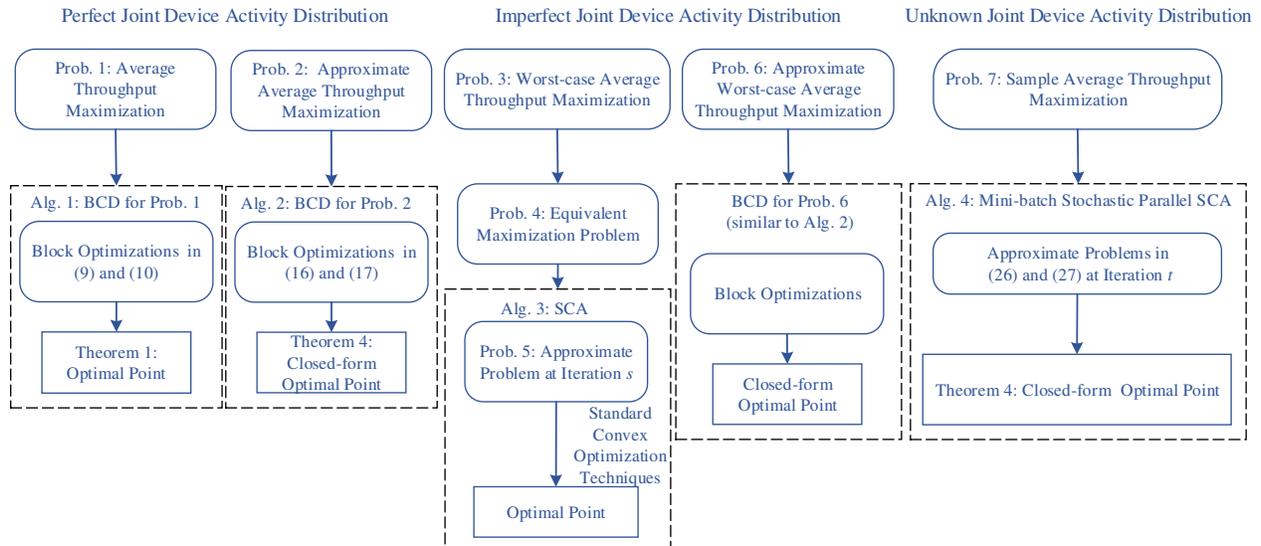}
	\vspace{-15mm}
	\caption{\small{Solution framework.}}\label{Q1figure}
	\vspace{-5mm}
\end{figure}

Motivated by Example 1, in Section III, Section IV, and Section V, we shall
optimize the
 preamble selection distributions $\bf A$ and access barring factor $\epsilon$ to
 maximize the average, worst-case average, and sample average throughputs for given $\mathbf p$, $\mathcal P$, and $\mathbf x_i$, $i \in \mathcal I$ in the cases of perfect, imperfect, and unknown joint device activity distributions, respectively,
 as shown in Fig.~\ref{Q1figure}.
 In what follows, we assume that the optimization problems are solved at the BS given knowledge of $\mathbf p$, $\mathcal P$, and $\mathbf x_i$, $i \in \mathcal I$, and the solutions are then sent to all devices for implementation.\footnote{In practice, $\mathbf p $ may change slowly over time.			
 		For practical implementation, we can re-optimize the preamble selection and access barring using the proposed methods 
 		when the change of $\mathbf p$ is sufficiently large, and the obtained stationary point for an outdated $\mathbf p$ generally serves as a good initial point for the latest $\mathbf p$.}

\section{Performance Optimization for Perfect Joint Device Activity Distribution}
In this section, we consider the average throughput maximization in the case of perfect joint device activity distribution.
First, we formulate the average throughput maximization problem, which
 is a challenging nonconvex problem.
Then, we develop an iterative algorithm to obtain a stationary point of the original problem.
Finally, we develop a low-complexity iterative algorithm to obtain a stationary point of an approximate problem.

\subsection{Problem Formulation}
In the case of perfect joint device activity distribution, we optimize the
 preamble selection distributions $\bf A$ and access barring factor $\epsilon$ to
 maximize the average throughput $ \bar T({\bf A},{\epsilon}, \bf p )$ in~(\ref{T_Avg.New}) subject to the constraints on~$({\bf A},\epsilon)$ in~(\ref{C1}), (\ref{C12}), (\ref{C2}), and (\ref{C3}).
\begin{Prob}[Average Throughput Maximization]\label{Prob:Perfectcase}
  \begin{align}
&\max_{\bf{A},\epsilon} \quad \bar T(\mathbf {A},\epsilon,\bf p)  \nonumber\\
&\ \mathrm{s.t.}
\quad \text{(\ref{C1})}, \ \text{(\ref{C12})}, \ \text{(\ref{C2})},\  \text{(\ref{C3})}. \nonumber
 \end{align}
 \end{Prob}

The objective function $\bar T(\mathbf {A},\epsilon,\bf p)$ is nonconcave in $({\bf A},\epsilon)$, and the constraints in~(\ref{C1}), (\ref{C12}),~(\ref{C2}), and~(\ref{C3}) are linear, corresponding to a convex feasible set.
Thus, Problem~\ref{Prob:Perfectcase} is nonconvex with a convex feasible set.
In general, a globally optimal point of a nonconvex problem cannot be obtained
effectively and efficiently.
Obtaining a stationary point is the classic goal for dealing with a nonconvex problem with a convex feasible set.

\subsection{Stationary Point }
We propose an iterative algorithm based on the BCD method~\cite{Bertsekas1998NP}, to obtain a stationary point of Problem~\ref{Prob:Perfectcase}.\footnote{One can obtain a stationary point of Problem~\ref{Prob:Perfectcase} using SCA~\cite{razaviyayn2014successive}. As the approximate convex optimization problem in each iteration has no analytical solution, SCA is not as efficient as the proposed one.}
Specifically, we divide the variables $\left({\bf A}, \epsilon\right)$
 into $K+1$ blocks, i.e., ${\bf a}_{k}, \ k \in \mathcal K $ and $ \epsilon $.
In each iteration of the proposed algorithm, all $K+1$ blocks are sequentially updated.
At each step of one iteration,
 we maximize $  \bar T(\mathbf {A},\epsilon,\bf p)  $
 with respect to one of the $K+1$ blocks.
{For ease of illustration, in the following, we also write
 $\bar T(\mathbf {A},\epsilon,\bf p)$ as $\bar T( {\bf a}_{k},{\bf a}_{-k} ,\epsilon,\bf p)$, where ${\bf a}_{-k}\triangleq({\bf a}_{\ell})_{\ell \in {\mathcal K}, \ell\neq k} $.}
Given ${\bf a}_{-k}$ and $\epsilon$ obtained in the previous step, the
 block coordinate optimization with respect to
${\bf a}_{k}$ is given by
  \begin{align}
  &\max_{{\bf a}_k} \quad
      \bar T\left( {\bf a}_k ,{\bf a}_{-k} ,  \epsilon,\bf p \right) , \quad  k\in \mathcal K \label{Prob:Perfectcase_RAO}\\
   &\ \mathrm{s.t.}   \quad   \text{(\ref{C2})}, \ \text{(\ref{C3})}. \nonumber
 \end{align}
Given $\bf A$ obtained in the previous step, the block coordinate optimization with respect to $\epsilon$ is given by
\begin{align}
  &\max_{\epsilon} \quad
     \bar T\left({\bf A}, \epsilon ,\bf p \right)
      \label{Prob:Perfectcase_ACB}  \\
   &  \ \mathrm{s.t.}  \quad   \text{(\ref{C1})}, \ \text{(\ref{C12})}.   \nonumber
 \end{align}
Each problem in (\ref{Prob:Perfectcase_RAO}) is a linear program (LP) with $N$
 variables and $N+1$ constraints.
The problem in (\ref{Prob:Perfectcase_ACB}) is a polynomial programming
 with a single variable and two constraints.

Next, we obtain optimal points of the problems in (\ref{Prob:Perfectcase_RAO}) and (\ref{Prob:Perfectcase_ACB}).
Define\footnote{Note that $ Q_{k,n}({\bf a}_{-k}, \epsilon,{\bf p}) =\frac{\partial \bar T ({\bf a}_k, {\bf a}_{-k},\epsilon,{\bf p})}{ \partial a_{k,n} } $, $ k\in \mathcal K$, $n\in \mathcal N$ and $ q({\bf A},\epsilon,{\bf p})  =\frac{\partial \bar T ({\bf A}, \epsilon,{\bf p})}{\partial \epsilon}  $.}
\begin{align}
&  Q_{k,n}({\bf a}_{-k}, \epsilon,{\bf p})
\triangleq \sum\limits_{m=1}^{K}
       m(-1)^{m-1}\epsilon^{m}\sum\limits_{
    \mathcal K' \subseteq {\mathcal K}: k\in \mathcal K', |\mathcal K' |=m}
   \left( \sum\limits_{{\bf x \in \mathcal X}}
p_{\bf x}
  \prod\limits_{\ell \in \mathcal K'  }{x}_{\ell}\right) \prod\limits_{\ell\in \mathcal K' : \ell \neq k  }  a_{\ell,n} ,   \label{Q_kn}
  \end{align}
\begin{align}
 & q({\bf A},\epsilon,{\bf p}) \triangleq
 \sum\limits_{m=1}^{K}
    m^2(-1)^{m-1} \epsilon^{m-1} \sum\limits_{n\in \mathcal N}\sum\limits_{
    \mathcal K' \subseteq {\mathcal K}:  |\mathcal K' |=m}
   \left( \sum\limits_{{\bf x \in \mathcal X}}
p_{\bf x}
  \prod\limits_{k \in \mathcal K'}{x}_{k}\right)  \prod\limits_{k \in \mathcal K'}  a_{k,n}   . \label{q}
\end{align}%
Denote $\mathcal B({\bf A},{\bf p}) \triangleq \{z \in [0,1]: q( {\bf A},z,{\bf p}) =0\}$ as the set of roots of equation
 $ q({\bf A},z,{\bf p})=0$ that lie in the interval $[0,1]$.
Based on the structural properties of the block coordinate optimization
 problems in (\ref{Prob:Perfectcase_RAO}) and (\ref{Prob:Perfectcase_ACB}), we can
 obtain their optimal points.
\begin{theorem}[Optimal Points of Problems
in (\ref{Prob:Perfectcase_RAO}) and (\ref{Prob:Perfectcase_ACB})]\label{Thm_OptimmalSolutionBCD}
A set of optimal points of each block coordinate optimization in~(\ref{Prob:Perfectcase_RAO}) is given by
\begin{align}
 \left\{ {\bf e}_{m} :     m \in  \mathop{\arg\max}_{n\in\mathcal N} \ Q_{k,n}({\bf a}_{-k}, \epsilon,{\bf p}) \right\}, \  k\in \mathcal K, \label{PerUpdate_a}
\end{align}
and a set of optimal points of the block coordinate optimization in~(\ref{Prob:Perfectcase_ACB}) is given by
\begin{align}
 \mathop{\arg\max}_{\epsilon\in {{\mathcal B}({\bf a},{\bf p})}\cup\{1\} }
  \bar T ( {\bf A},\epsilon,{\bf p}) .\label{PerUpdate_epsilon}
\end{align}

\end{theorem}

\begin{IEEEproof}
 Please refer to Appendix B.
 \end{IEEEproof}
The optimal point in (\ref{PerUpdate_a}) indicates that each device $k $ selects the preamble corresponding to the maximum average throughput (increase rate of the average throughput) conditioned on selected by device $k$ for given $ (\mathbf a_{-k}, \epsilon)$.
  The overall computational complexity for determining the sets in (\ref{PerUpdate_a}) is $\mathcal O (NK^22^K) $, and the overall computational complexity for determining the set in (\ref{PerUpdate_epsilon}) is $\mathcal O(  NK^22^K  ) $.
 The detailed complexity analysis can be found in Appendix C.
 As $\bf A$ is usually sparse during the iterations, the actual computational complexities for obtaining~(\ref{PerUpdate_a}) and~(\ref{PerUpdate_epsilon}) are much lower.

Finally, the details of the proposed iterative algorithm are summarized in Algorithm~\ref{Alg_PBCD}.
Specifically, in Steps~$4-7$, $\mathbf a_k$, $k \in \mathcal K$ are updated one by one;
in Steps~$9-11$, $\epsilon$ is updated.
Step~$5$ and Step~$9$ are to ensure the convergence of Algorithm~\ref{Alg_PBCD} to a stationary point.
Based on the proof for~\cite[Proposition 2.7.1]{Bertsekas1998NP}, we can show the following result.

\begin{theorem}[Convergence of Algorithm~\ref{Alg_PBCD}]\label{Thm_BCD}
		Algorithm \ref{Alg_PBCD} returns a stationary point of Problem~\ref{Prob:Perfectcase} in a finite number of iterations.
\end{theorem}

\begin{IEEEproof}
 Please refer to Appendix D.
 \end{IEEEproof}


In practice, we can run Algorithm~\ref{Alg_PBCD} multiple
times (with possibly different random initial points) to obtain multiple
stationary points and choose the stationary point with the largest
objective value as a suboptimal point of Problem~\ref{Prob:Perfectcase}.\footnote{Generally speaking, finding a good stationary point involves numerical experiments and is more art than technology~\cite{Boyd2004convex}.
	However, we find that $  ((\mathbf e_{n_k})_{k \in \mathcal K},  1 )$ is generally a good initial point that yields a faster convergence speed with numerical experiments.
	The intuition is that when $(\mathbf A, \epsilon) = ((\mathbf e_{n_k})_{k \in \mathcal K},  1 )$, the contending devices for each preamble are statistically balanced irrespective of $\mathbf p$.}
The average throughput of the best obtained stationary point and the computational complexity increase
	with the number of times that Algorithm~\ref{Alg_PBCD} is run.
We can choose a suitable number to balance the increases of the average throughput and computational complexity.

Based on Algorithm~\ref{Alg_PBCD} and Theorem~\ref{Thm_BCD}, we can characterize an optimality property of a globally optimal point of Problem~\ref{Prob:Perfectcase}.
 \begin{theorem}[Optimality Property] \label{Lem_Prop_Opt}
	There exists at least one globally optimal point $({\bf A}^{*},\epsilon^{*})$ of Problem \ref{Prob:Perfectcase}, which satisfies
	${\bf a}_{k}^{*} ={\bf e}_{n_k},  k\in \mathcal K$, for some $n_k \in \mathcal N$, $k\in\mathcal K$.
\end{theorem}

\begin{IEEEproof}
	Please refer to Appendix E.
\end{IEEEproof}

\begin{algorithm}[t]
	\caption{Obtaining A Stationary Point of Problem~\ref{Prob:Perfectcase}}
	\label{Alg_PBCD}
	\begin{multicols}{2}
		\begin{algorithmic}[1]
			\small{\STATE \textbf{initialization:} 		
				for $k \in \mathcal K$, set ${\bf a}_k := {\bf e}_{n_k}$, where $n_k$ is uniformly chosen from $\mathcal N$ at random, and set $\epsilon:= 1$.
				\STATE \textbf{repeat}
				\STATE  ${\bf A}_{\text{last}} :=\bf A$.
				\STATE \textbf{for $k \in \mathcal K$ do}
				\STATE \quad \textbf{if} ${\bf a}_k$ does not belong to the set in (\ref{PerUpdate_a})
				\STATE \quad \quad ${\bf a}_k $ is randomly chosen from the set in (\ref{PerUpdate_a}).
				\STATE \quad\textbf{end if}
				\STATE  \textbf{end for}
				\STATE  \textbf{if} $\epsilon$ does not belong to the set in (\ref{PerUpdate_epsilon})
				\STATE  \quad $\epsilon$ is randomly chosen from the set in (\ref{PerUpdate_epsilon}).
				\STATE  \textbf{end if}
				\STATE \textbf{until} $  {\bf A}_{\text{last}}=  {\bf A }$.
				\normalsize }
		\end{algorithmic}
	\end{multicols}
\end{algorithm}

Theorem \ref{Lem_Prop_Opt} indicates that there exists a deterministic preamble
selection rule that can achieve the maximum average throughput.
It is worth noting that the proposed stationary point satisfies the optimality property in Theorem~\ref{Lem_Prop_Opt}.
In Section III-C, we shall see that the low-complexity solution also satisfies this optimality property.


\subsection{Low-complexity Solution }
From the complexity analysis for obtaining a stationary point of Problem~\ref{Prob:Perfectcase}, we know that
 Algorithm~\ref{Alg_PBCD} is computationally expensive when $K$ or $N$ is large.
In this part, we develop a low-complexity algorithm, which is applicable for large $K$ or $N$, to obtain a stationary point of an approximate problem of Problem~\ref{Prob:Perfectcase}.
Later, in Section VI, we shall show that the performance of the low-complexity algorithm is comparable with Algorithm~\ref{Alg_PBCD}.

Motivated by the approximations of $\bar T(\mathbf {A},\epsilon,{\bf p})$ in \cite{Popovski18SPAWC},
	we approximate the complicated function $\bar T(\mathbf {A},\epsilon,{\bf p})$, which has $N2^K$ terms, with a simpler function
\begin{align}
	\tilde T(\mathbf {A},\epsilon,{\bf p}) & \triangleq
 \epsilon \sum\limits_{k\in \mathcal {K}} \sum\limits_{{\bf x} \in \mathcal X }p_{\bf x}x_k
	-\epsilon^2 \sum\limits_{n\in \mathcal {N}} \sum\limits_{k\in \mathcal {K }}a_{k,n} \sum\limits_{\ell  \in {\mathcal K}: \ell >k}a_{\ell ,n}
	\sum\limits_{{\bf x} \in \mathcal X }p_{\bf x} x_kx_{\ell} ,
	\label{T_LB}
\end{align}
 which has $ 1+ \frac{K(K-1)}{2}$ terms.
 The detailed reason for the approximation can be found in~\cite{LWarxiv}.
Note that $  \sum\nolimits_{{\bf x} \in \mathcal X }p_{\bf x}x_k  $ and $  \sum\nolimits_{{\bf x} \in \mathcal X }p_{\bf x}x_kx_{\ell}  $ ($k<\ell$) represent the probability of device $k$ being active and the probability of devices $k$ and $\ell$ being active, respectively.
By comparing~(\ref{T_LB}) with~(\ref{T_Avg.New}), we can see that $ \tilde T (\mathbf {A},\epsilon,{\bf p})  $ captures the active probabilities of every single device and every two devices.
Accordingly, we consider the following approximate problem of Problem~\ref{Prob:Perfectcase}.

\begin{Prob}
[Approximate Average Throughput Maximization]
\label{Prob:Perfectcase_LB}
  \begin{align}
&{\max_{\bf{A},\epsilon}} \quad  \tilde T\left(\mathbf {A},\epsilon,{\bf p}\right) \nonumber\\
 & \ \mathrm{s.t.} \quad \text{(\ref{C1})},\  \text{(\ref{C12})},\ \text{(\ref{C2})},  \ \text{(\ref{C3})} .\nonumber
 \end{align}
 \end{Prob}

Analogously, using the BCD method, we propose a computationally efficient iterative algorithm, with more performance guarantee than the heuristic method in \cite{Popovski18SPAWC}, to obtain a stationary point of Problem \ref{Prob:Perfectcase_LB}.
Specifically, variables $\left({\bf A}, \epsilon\right)$ are divided into $K+1$ blocks, i.e., $ {\bf a}_{k}, \ k \in \mathcal K $ and $ \epsilon $.
For ease of illustration, we also write $\tilde T(\mathbf {A},\epsilon,\bf p)$ as $\tilde T( {\bf a}_{k},{\bf a}_{-k} ,\epsilon,\bf p)$ in the sequel.
Given ${\bf a}_{-k}$ and $\epsilon$ obtained in the previous step, the block coordinate optimization with respect to ${\bf a}_{k}$ is
given by
  \begin{align}
  &\max_{{\bf a}_k} \quad
    \tilde T\left( {\bf a}_k ,{\bf a}_{-k} ,   \epsilon,{\bf p}\right)
        ,\quad k \in \mathcal K  \label{Prob:Perfectcase_RAOLB}\\
   &\ \mathrm{s.t.}   \quad   \text{(\ref{C2})}, \ \text{(\ref{C3})}.  \nonumber
 \end{align}
Given $\bf A$ obtained in the previous step, the block coordinate optimization with respect to $\epsilon$ is given by
\begin{align}
  &\max_{\epsilon} \quad
    \tilde T  \left({\bf A},\epsilon ,{\bf p}  \right)
      \label{Prob:Perfectcase_ACBLB}  \\
   &  \ \mathrm{s.t.}  \quad \      \text{(\ref{C1})}, \ \text{(\ref{C12})}. \nonumber
 \end{align}
Each problem in (\ref{Prob:Perfectcase_RAOLB}) is an LP with $N$ variables and $N+1$ constraints, and the problem in (\ref{Prob:Perfectcase_ACBLB}) is a quadratic program (QP) with a single variable and two constraints.
 It is clear that the convex problems in (\ref{Prob:Perfectcase_RAOLB}) and (\ref{Prob:Perfectcase_ACBLB}) are much simpler than those in
 (\ref{Prob:Perfectcase_RAO}) and (\ref{Prob:Perfectcase_ACB}), respectively.
Based on the structural properties of the block coordinate optimization problems in (\ref{Prob:Perfectcase_RAOLB}) and (\ref{Prob:Perfectcase_ACBLB}), we can obtain their optimal points.

\begin{theorem}[Optimal Points of Problems in (\ref{Prob:Perfectcase_RAOLB}) and (\ref{Prob:Perfectcase_ACBLB})]\label{Thm_OptimmalSolutionBCDLB}
A set of optimal points of each block coordinate optimization in (\ref{Prob:Perfectcase_RAOLB}) is given by
\begin{align}
 \left\{ {\bf e}_{m} :     m \in  \mathop{\arg\min}_{n\in\mathcal N}
   \quad
     \sum\limits_{\ell \in {\mathcal K}: \ell \neq k}   a_{\ell,n}\sum\limits_{{\bf x} \in \mathcal X }p_{\bf x} x_k x_{\ell} \right\}
      ,\ k\in\mathcal K,  \label{PerLBUpdate_a}
\end{align}
and the optimal point of the block coordinate optimization in~(\ref{Prob:Perfectcase_ACBLB}) is given by
\begin{align}
\min\left(1,  \frac{\sum\limits_{k\in \mathcal {K}}
 \sum\limits_{{\bf x} \in \mathcal X }p_{\bf x}x_k}
 {
  2\sum\limits_{n\in \mathcal {N}} \sum\limits_{k\in \mathcal {K}}a_{k,n}\sum\limits_{\ell \in {\mathcal K}:\ell> k} a_{\ell,n} \sum\limits_{{\bf x} \in \mathcal X }p_{\bf x} x_kx_{\ell}   }\right ).\label{PerLBUpdate_epsilon}
\end{align}

\end{theorem}

\begin{IEEEproof}
Theorem \ref{Thm_OptimmalSolutionBCDLB} can be proved in a similar way to Theorem \ref{Thm_OptimmalSolutionBCD}.
 \end{IEEEproof}

The optimal point in (\ref{PerLBUpdate_a}) indicates that each device $k$ selects the preamble with the minimum number of contending devices conditioned on selected by device $k$ for given $ (\mathbf a_{-k}, \epsilon)$.
In the following, we analyze the computational complexity for solving the problems in~(\ref{Prob:Perfectcase_RAOLB}) and~(\ref{Prob:Perfectcase_ACBLB}) according to Theorem~\ref{Thm_OptimmalSolutionBCDLB}.
  As constants $\sum\nolimits_{k\in \mathcal {K}}
  \sum\nolimits_{{\bf x} \in \mathcal X }p_{\bf x}x_k$ and $\sum\nolimits_{{\bf x} \in \mathcal X }p_{\bf x}x_{\ell}x_k $, $k,\ell \in \mathcal K,k<\ell $ are computed in advance,
the corresponding computational complexities are not considered below.
For all $k \in \mathcal K$ and $n \in \mathcal N$, the computational complexity for
 calculating $\sum\nolimits_{\ell \in {\mathcal K}: \ell \neq k}   a_{\ell,n}\sum\nolimits_{{\bf x} \in \mathcal X } p_{\bf x} x_k x_{\ell}$ is $\mathcal O ( K)$.
Furthermore, for all $k \in \mathcal K$, the computational complexity of finding the largest one among $ \sum\nolimits_{\ell \in {\mathcal K}: \ell \neq k}   a_{\ell,n}\sum\nolimits_{{\bf x} \in \mathcal X }p_{\bf x} x_k x_{\ell}$, $n\in \mathcal N$ is $\mathcal O (N)$.
Thus, the overall computational complexity for determining the sets in (\ref{PerLBUpdate_a}) is $\mathcal O ( NK^2) +\mathcal O ( NK ) = \mathcal O ( NK^2) $.
Analogously, the computational complexity for obtaining the closed-form optimal point in (\ref{PerLBUpdate_epsilon}) is $\mathcal O(NK^2)$.
It is obvious that the computational complexities for obtaining the optimal points given by Theorem~\ref{Thm_OptimmalSolutionBCDLB} are much lower than those for obtaining the optimal points given by Theorem~\ref{Thm_OptimmalSolutionBCD}.
Furthermore, it is worth noting that the optimal points
given by Theorem~\ref{Thm_OptimmalSolutionBCDLB} do not rely on the active probabilities of more than two devices.

Finally, the details of the proposed iterative algorithm are summarized in Algorithm~\ref{Alg_PLBBCD}.
Specifically, in Steps $4-7$, $\mathbf a_k$, $k \in \mathcal K$ are updated one by one;
	in Steps~$9-11$, $\epsilon$ is updated.
Step~$5$ and Step $9$ are to ensure the convergence of Algorithm~\ref{Alg_PLBBCD} to a stationary point.
Similarly, we have the following results.
\begin{theorem}[Convergence of Algorithm~\ref{Alg_PLBBCD}]\label{Thm_LBBCD}
Algorithm \ref{Alg_PLBBCD} returns a stationary point of Problem~\ref{Prob:Perfectcase_LB} in a finite number of iterations.
%
\end{theorem}

\begin{IEEEproof}
Theorem \ref{Thm_LBBCD} can be proved in a similar way to Theorem \ref{Thm_BCD}.
 \end{IEEEproof}

\begin{algorithm}[t]
\caption{Obtaining A Stationary Point of Problem \ref{Prob:Perfectcase_LB}}
\label{Alg_PLBBCD}
\begin{multicols}{2}
\begin{algorithmic}[1]
\small{ \STATE \textbf{initialization:}
 for $k \in \mathcal K$, set ${\bf a}_k := {\bf e}_{n_k}$, where $n_k$ is uniformly chosen from $\mathcal N$ at random, and set $\epsilon:= 1$.
\STATE \textbf{repeat}
\STATE ${\bf A}_{\text{last}}:=\bf A$.
\STATE \textbf{for $k \in \mathcal K$ do}
\STATE \quad \textbf{if} ${\bf a}_k$ does not belong to the set in  (\ref{PerLBUpdate_a})
\STATE \quad \quad ${\bf a}_k $ is randomly chosen from the set in  (\ref{PerLBUpdate_a}).
\STATE \quad\textbf{end if}
\STATE  \textbf{end for}
\STATE  \textbf{if} $\epsilon$ does not belong to the set in (\ref{PerLBUpdate_epsilon})
\STATE  \quad $\epsilon$ is randomly chosen from the set in  (\ref{PerLBUpdate_epsilon}).
\STATE  \textbf{end if}
\STATE  \textbf{until} $  {\bf A}_{\text{last}}=  {\bf A }$.
\normalsize}
\end{algorithmic}
\end{multicols}
\end{algorithm}

Analogously, we can run Algorithm~\ref{Alg_PLBBCD} multiple
	times (with possibly different random initial points) to obtain multiple
	stationary points of Problem~\ref{Prob:Perfectcase_LB} and choose the stationary point with the largest average throughput as a low-complexity suboptimal point of Problem~\ref{Prob:Perfectcase}.

\section{Robust Optimization for Imperfect Joint Device Activity Distribution}

In this section, we consider the worst-case average throughput maximization in the case of imperfect joint device activity distribution.
First, we formulate the worst-case average throughput maximization problem, which is a challenging max-min problem.
Then, we develop an iterative algorithm to obtain a KKT point of an equivalent problem.
Finally, we develop a low-complexity iterative algorithm to obtain a stationary point of an approximate problem.

 \subsection{Problem Formulation }
 In the case of imperfect joint device activity distribution, we optimize the
 preamble selection distributions $\bf A$ and access barring factor $\epsilon$ to
 maximize the worst average throughput $ \bar T_{\text{wt}}({\bf A},{\epsilon}, \bf p )$ in (\ref{T_wt}) subject to the constraints on $({\bf A},\epsilon)$ in (\ref{C1}), (\ref{C12}), (\ref{C2}), and (\ref{C3}).
\begin{Prob}[Worst-case Average Throughput Maximization]
\label{Prob:Imperfectcase}
   \begin{align}
 \max\limits_{{\bf A}, \epsilon} &\quad \bar T_{\text{wt}}({\bf A},{\epsilon},\mathcal P)  \nonumber \\
\mathrm{s.t.} & \quad \text{(\ref{C1})}, \ \text{(\ref{C12})}, \ \text{(\ref{C2})},  \ \text{(\ref{C3})}. \nonumber
 \end{align}
 \end{Prob}

 Note that we explicitly consider the
estimation error of the joint device activity distribution in the optimization.
The objective function $\bar T_{\text{wt}}({\bf A},{\epsilon},\mathcal P)$ is nonconcave in $({\bf A},\epsilon)$, and the constraints in (\ref{C1}), (\ref{C12}), (\ref{C2}), and (\ref{C3}) are linear.
Thus, Problem \ref{Prob:Imperfectcase} is nonconvex.
Moreover, note that the objective function $\bar T_{\text{wt}}({\bf A},{\epsilon},\mathcal P) \triangleq \min_{{\bf p} \in \mathcal P} \bar T({\bf A},{\epsilon},{\bf p} )$ does not have an analytical form.

  \subsection{KKT Point}
 Problem~\ref{Prob:Imperfectcase} is a challenging max-min problem.
 In this part, we solve Problem~\ref{Prob:Imperfectcase} in two steps~\cite{Ye2019}.
First, we transform the max-min problem in Problem \ref{Prob:Imperfectcase} to an equivalent maximization problem.
As the inner problem $ \min_{{\bf p} \in \mathcal P} \bar T({\bf A},\epsilon,{\bf p})$ is an LP with respect to $\bf p$ and strong duality holds for the LP, the inner problem $ \min_{{\bf P} \in \mathcal P} \bar T({\bf A},\epsilon,{\bf p})$ shares the same optimal value as its dual problem.
Furthermore, to facilitate algorithm design,\footnote{In the constraints in (\ref{P5constrains}), variables $\bf B$, $\epsilon$ ${\boldsymbol{\lambda}}$, and $\nu$ are coupled.
	Thus, we cannot apply the BCD method to solve Problem~\ref{Prob:ImperfectcaseEQP}.}
we can transform Problem~\ref{Prob:Imperfectcase} to the following equivalent problem by a change of variables ${\bf B }=  \epsilon \bf A$, where ${\bf B} \triangleq (b_{k,n})_{k \in \mathcal {K}, n \in \mathcal N}$.

\begin{Prob}[Equivalent Problem of Problem \ref{Prob:Imperfectcase}]
\label{Prob:ImperfectcaseEQP}
     \begin{subequations}\label{Prob_ImperfectcaseEQP}
      \begin{align}
\max\limits_{{\bf B},{\boldsymbol{\lambda}}\succeq0,\epsilon  ,\nu }
 &\quad  \left(  \sum\limits_{{\bf x}\in \mathcal X  }\underline{p}_{\bf x}  - 1\right)\nu +
 \sum\limits_{{\bf x}\in \mathcal X  } \left( (\underline{p}_{\bf x} - \overline{p}_{\bf x})\lambda_{\bf x}
 +\underline{p}_{\bf x}  T({\bf B},1, {\bf x})  \right)
 \nonumber \\
\mathrm{s.t.}&  \quad\  \ \text{(\ref{C1})} , \ \text{(\ref{C12})}, \nonumber \\
&\quad \ b_{k,n} \geq 0, \ k \in \mathcal {K}, n \in \mathcal N, \label{Cbb1}\\
 &\quad \ \sum\limits_{n \in \mathcal {N}}{{b_{k,n}}}=\epsilon,\ k \in \mathcal {K} ,\label{Cbb2} \\
& \quad \   \nu +\lambda_{\bf x}+  T({\bf B},1, {\bf x}) \geq0 ,
     \ {\bf x} \in \mathcal X,
   \label{P5constrains}
 \end{align}
\end{subequations}
 \end{Prob}
 where $\boldsymbol{\lambda} \triangleq (\lambda_{\bf x})_{{\bf x}\in \mathcal X} $,
  and $ T({\bf A},\epsilon, {\bf x})$ is given by (\ref{T_x}).
Let $({\bf B}^{\star},{\boldsymbol{\lambda}}^{\star}, \epsilon^{\star}, \nu^{\star} )$ denote an optimal point of Problem \ref{Prob:ImperfectcaseEQP}.

The equivalence between Problem \ref{Prob:Imperfectcase} and Problem \ref{Prob:ImperfectcaseEQP} is summarized below.\footnote{As $p_{\bf 0} < 1$, we can easily show $\epsilon^{\star}> 0  $.}
\begin{lemma}[Equivalence between Problem \ref{Prob:Imperfectcase} and Problem \ref{Prob:ImperfectcaseEQP}] \label{Lem_ImpefectHL_Equi}
$ (\frac{{\bf B}^{\star}}{\epsilon^{\star}},  \epsilon^{\star})$ is an optimal point of Problem \ref{Prob:Imperfectcase}.
\end{lemma}

\begin{IEEEproof}
 Please refer to Appendix F.
 \end{IEEEproof}

Next, based on Lemma~\ref{Lem_ImpefectHL_Equi}, we can solve Problem \ref{Prob:ImperfectcaseEQP} instead of Problem~\ref{Prob:Imperfectcase}.
Problem~\ref{Prob:ImperfectcaseEQP} is nonconvex with a nonconvex feasible set, as $   T({\bf B},1, {\bf x}) $, $\mathbf x \in \mathcal X$ in the objective function and the constraint functions in~(\ref{P5constrains}) are nonconcave.
Note that obtaining a KKT point is the classic goal for dealing with a nonconvex problem with a nonconvex feasible set.
In what follows, we propose an iterative algorithm to obtain a KKT point of Problem~\ref{Prob:ImperfectcaseEQP} using SCA.
Specifically, at iteration~$s$, we update $({\bf B}^{(s) },{\boldsymbol{\lambda}}^{ (s) },\epsilon^{ (s) },\nu^{(s) } )$
 by solving an approximate convex problem parameterized by
 ${\bf B}^{(s-1)}$
obtained at iteration $s-1$.
For notation convenience, define\footnote{Note that $\frac{\partial  T({{\bf A}},\epsilon, {\bf x}) }{\partial a_{k,n}}   =  x_k\epsilon	g_{k,n}( {\bf A}, \epsilon,{\bf x}) $ , $ k \in \mathcal K, n \in \mathcal N$ and $\frac{\partial  T({{\bf A}},\epsilon, {\bf x}) }{\partial \epsilon}   = \sum\nolimits_{k \in \mathcal K} x_k \sum\nolimits_{n \in \mathcal N} a_{k,n}	g_{k,n}( {\bf A}, \epsilon,{\bf x})$.}
\begin{align}
	g_{k,n}( {\bf A} , \epsilon,{\bf x}) \triangleq  \prod\limits_{\ell \in \mathcal K:\ell\neq k}
	\!\!\!\!
	(1 \!\! - \!x_{\ell}a_{\ell,n}\epsilon)
	-
	\epsilon  \!\!\!\sum\limits_{\ell\in\mathcal K:\ell\neq k}
	\!\!\!\!\!
	x_{\ell}a_{\ell,n}
	\!\!\!\!\!\!\!
	\prod\limits_{j \in \mathcal K:j\neq \ell,j\neq k}
	\!\!\!\!\!\!\!\!
	(1-x_{j}a_{j,n}\epsilon) , \   k \in \mathcal K, n \in \mathcal N. \label{g_{k,n}}
\end{align}
We choose
 \begin{align}
 \tilde h({\bf B},{\boldsymbol{\lambda}},{\nu},{\bf B}^{ (s-1)  } ) \triangleq&
   \left(  \sum\limits_{{\bf x}\in \mathcal X  } \underline{p}_{\bf x} -1 \right )\nu + \sum\limits_{{\bf x}\in \mathcal X  } \left( (\underline{p}_{\bf x} - \overline{p}_{\bf x})\lambda_{\bf x}
 +\underline{p}_{\bf x} T({{\bf B}^{ (s-1)  }},1, {\bf x}) \right)   \nonumber
    \end{align}
\begin{align}
 &+
  	\sum\limits_{k\in\mathcal K}
  \sum\limits_{n\in\mathcal N}\sum\limits_{{\bf x}\in \mathcal X }  \underline{p}_{\bf x} x_k
   	g_{k,n}( {\bf B}^{(s-1)}, 1,{\bf x})
  \left(b_{k,n}-b_{k,n}^{ (s-1) } \right)
   \nonumber  \\
& -\frac{ 1}{2} \sqrt{N  \sum\limits_{k \in \mathcal K} \sum\limits_{\ell \in \mathcal K: \ell\neq k } \big(\sum\limits_{\mathbf x \in \mathcal X} \underline p_{\mathbf x} x_k x_{\ell} \| \mathbf x\|_1 \big)^2}  \sum\limits_{k\in\mathcal K}\sum\limits_{n\in\mathcal N}   \left(b_{k,n}-b_{k,n}^{ (s-1)  }\right)^2 ,
  \label{Approximation_Imperfectcase_h}
 \end{align}
 where $\|\cdot\|_{1}$ denotes the $\ell_1$-norm, as an approximate function of the objective function in Problem~\ref{Prob:ImperfectcaseEQP} at iteration $s$.
 In addition, we choose
\begin{align}
 \tilde h_\text{c}({\bf B},{\boldsymbol{\lambda}},{\nu},{\bf B}^{ (s-1) }, {\bf x} )
\triangleq
&
  \nu +\lambda_{\bf x} + T\left({\bf B}^{  (s-1)  },1, {\bf x} \right)
  +
  \sum\limits_{k\in\mathcal K}
  \sum\limits_{n\in\mathcal N}
   	 x_kg_{k,n}( {\bf B}^{(s-1)}, 1,{\bf x})  \left(b_{k,n}-b_{k,n}^{ (s-1)  }\right)
  \nonumber \\
&-  \frac{ 1 }{2} \sqrt{N    } \| \mathbf x\|_1^2\sum\limits_{k\in\mathcal K} x_{k} \sum\limits_{n\in\mathcal N}
   \left(b_{k,n}-b_{k,n}^{(s-1)} \right)^2  
    \label{Approximation_Imperfectcase_hx}
\end{align}
as an approximate function of the constraint function for ${\bf x} \in \mathcal X$ in~(\ref{P5constrains}) at iteration $s$.
Note that the concave components of the objective function and the constraint functions in Problem~\ref{Prob:ImperfectcaseEQP} are left unchanged, and the other nonconcave components, i.e., $ \sum\nolimits_{{\bf x}\in \mathcal X} \underline p_{\bf x} T({\bf B},1, {\bf x}) $, in the objective function and $T({\bf B},1, {\bf x}), {\bf x} \in \mathcal X$ in~(\ref{P5constrains}) are minorized at $({\bf B},{\boldsymbol{\lambda}},\epsilon,  {\nu})=
  ({\bf B}^{(s-1)},{\boldsymbol{\lambda}}^{(s-1)},\epsilon^{(s-1)},  \nu^{(s-1)}) $,
  using the concave components based on their second-order Taylor expansions~\cite{SunMM}.
Then, at iteration $s$, we approximate Problem~\ref{Prob:ImperfectcaseEQP} with the following convex problem.

\begin{Prob}[Approximate Problem of Problem \ref{Prob:ImperfectcaseEQP} at Iteration $s$]
\label{Prob:ImperfectcaseEQPSCA}
  \begin{align}
    (\hat{\bf B}^{ (s) },\hat{\boldsymbol{\lambda}}^{ (s) },\hat\epsilon^{ (s) },\hat\nu^{ (s) } )
    \triangleq
    \mathop{\arg\max}\limits_{{\bf B},{\boldsymbol{\lambda}}\succ0,\epsilon,\nu  }
    & \quad
    \tilde h({\bf B},{\boldsymbol{\lambda}},  {\nu}  ,{\bf B}^{ (s-1) } )       \nonumber \\
 \mathrm{s.t.}&  \quad \ \text{(\ref{C1})}, \ \text{(\ref{C12})},
    \ \text{(\ref{Cbb1})}, \ \text{(\ref{Cbb2})}, \nonumber \\
 &   \quad  \tilde h_\text{c} ({\bf B},{\boldsymbol{\lambda}},{\nu}, {\bf B}^{ (s-1) } , {\bf x})  \geq 0,   \ {\bf x} \in \mathcal X.  \nonumber 
  \end{align}
  \end{Prob}


First, we analyze the computational complexity for determining the updated objective function and constraint functions of Problem~\ref{Prob:ImperfectcaseEQPSCA}.
The computational complexity for calculating $ x_k g_{k,n}( {\bf B}^{(s-1)}, 1,{\bf x})$, $k\in \mathcal K, n \in \mathcal N, {\bf x} \in \mathcal X$ is $\mathcal O ( NK^3 2^K)$.
Given $ x_k g_{k,n}( {\bf B}^{(s-1)}, 1,{\bf x})$, the computational complexity for calculating $ \sum\nolimits_{\mathbf x \in \mathcal X}\underline{p}_{\mathbf x} x_k g_{k,n}( {\bf B}^{(s-1)}, 1,{\bf x}) $ is $\mathcal O (NK2^K  )$.
Note that constants $\| \mathbf x\|_1^2$, $\mathbf x \in \mathcal X$ and $\sum\nolimits_{k \in \mathcal K} \sum\nolimits_{\ell \in \mathcal K: \ell\neq k } \big(\sum\nolimits_{\mathbf x \in \mathcal X} \underline p_{\mathbf x} x_k x_{\ell} \| \mathbf x\|_1 \big)^2$ are computed in advance.
Thus, at iteration $s$, the computational complexity for determining the updated objective function and constraint functions of Problem~\ref{Prob:ImperfectcaseEQPSCA} is $ \mathcal O(  NK^3 2^K)  +\mathcal O ( NK2^K)   =\mathcal O(   NK^3 2^K) $.
Then, we solve Problem~\ref{Prob:ImperfectcaseEQPSCA}.
Problem~\ref{Prob:ImperfectcaseEQPSCA} has $2+NK+2^K$ variables and $2+ N(K+1) +2^{(K+1)}$ constraints.
Thus, solving Problem~\ref{Prob:ImperfectcaseEQPSCA} by using an interior-point method has computational complexity $ \mathcal O \left( ( NK+2^K )^3 \right) $\cite[pp. 8]{Boyd2004convex}.
After solving Problem~\ref{Prob:ImperfectcaseEQPSCA},
we update
\begin{subequations}\label{ROSCAupdate}
\begin{align}
	&{\bf B}^{ (s)  } = (1-\gamma){\bf B}^{ (s-1) }+\gamma \hat{\bf B}^{ (s)}, \label{ROSCAupdate1}
	\ {\boldsymbol{\lambda}}^{  (s) } = (1-\gamma)\boldsymbol{\lambda}^{  (s-1)   }+ \gamma\hat{\boldsymbol{\lambda}}^{  (s) },  \\
	&\epsilon^{ (s)  }  = (1-\gamma) \epsilon^{ (s-1) } + \gamma\hat\epsilon^{(s)},
	\ \nu^{ (s) } = (1-\gamma) \nu^{ (s-1) } + \gamma\hat\nu^{(s)},\label{ROSCAupdate4}
\end{align}
\end{subequations}
where $\gamma \in (0,1]$ is a positive constant.

Finally, the details of the proposed iterative algorithm are summarized in Algorithm \ref{Alg_ROpt}.
Based on \cite[Theorem 1]{razaviyayn2014successive}, we can show the following result.

\begin{theorem}[Convergence of Algorithm \ref{Alg_ROpt}]\label{Thm_Ropt}
Let $({\bf B}^{\dag},\boldsymbol{\lambda}^{\dag},\epsilon^{\dag},\nu^{\dag} )$ be a limit point of the iterates generated by Algorithm \ref{Alg_ROpt}.
If the interior of the set $\{ ({\bf B},{\boldsymbol{\lambda}},\epsilon,\nu ) |   \ \tilde h_\text{c} ({\bf B},{\boldsymbol{\lambda}},{\nu}, {\bf B}^{ (\dag) } , {\bf x})  \geq 0,   \ {\bf x} \in \mathcal X , \text{(\ref{C1})}, \text{(\ref{C12})}, \text{(\ref{Cbb1})}, \text{(\ref{Cbb2})}\}$
is nonempty, then  $({\bf B}^{\dag},\boldsymbol{\lambda}^{\dag},\epsilon^{\dag},\nu^{\dag} )$ is a KKT point of Problem \ref{Prob:ImperfectcaseEQP}.
\end{theorem}

\begin{IEEEproof}
 Please refer to Appendix G.
 \end{IEEEproof}

 \begin{algorithm}[t]
\caption{Obtaining A KKT Point of
Problem~\ref{Prob:ImperfectcaseEQP}}
\label{Alg_ROpt}
\begin{multicols}{2}
\begin{algorithmic}[1]
\small{ \STATE \textbf{initialization:} Set 
$ ({\bf B}^{(0)},{\boldsymbol{\lambda}}^{(0)},\epsilon^{(0)}  ,\nu^{(0)} )$ $:=((\mathbf e_{n_k})_{k \in \mathcal K},\mathbf 1,  1,0)$, where $n_k$ is uniformly chosen from $\mathcal N$ at random, for all $k \in \mathcal K$, set $s := 0$, and choose $\mu>0$.
\STATE \textbf{repeat}
\STATE \quad Set $s:=s+1$.
\STATE \quad Calculate $ (\hat{\bf B}^{(s)} ,
{\hat {\boldsymbol{\lambda}}  }^{(s)},\hat \epsilon ^{(s)} \hat\nu^{(s)} )$ by solving Problem~\ref{Prob:ImperfectcaseEQPSCA}.
\STATE \quad Update $({\bf B}^{(s)},{\boldsymbol{\lambda}}^{(s)},\epsilon^{(s)} ,\nu^{(s)})$ according to (\ref{ROSCAupdate}).
\STATE \textbf{until} $ \| {\bf B}^{(s)} -{\bf B}^{(s-1)}    \|_{\mathbb F} \leq \mu $.
\normalsize}
\end{algorithmic}
\end{multicols}
\end{algorithm}

Analogously, we can run Algorithm~\ref{Alg_ROpt} multiple
times (with possibly different random initial points) to obtain multiple
stationary points of Problem~\ref{Prob:ImperfectcaseEQP} and choose the KKT point with the largest worst-case average throughput as a suboptimal point of Problem~\ref{Prob:ImperfectcaseEQP}, which can also be regarded as a suboptimal point of Problem~\ref{Prob:Imperfectcase} according to Lemma~\ref{Lem_ImpefectHL_Equi}.

\subsection{Low-complexity Solution }
From the complexity analysis for solving Problem~\ref{Prob:ImperfectcaseEQPSCA}, we know that
 Algorithm~\ref{Alg_ROpt} is computationally expensive when $K$ or $N$ is large.
In this part, we develop a low-complexity iterative algorithm, which is applicable for large $K$ or $N$, to obtain a stationary point of an approximate problem of Problem~\ref{Prob:Imperfectcase}.
Later, in Section VI, we shall show that this low-complexity algorithm can achieve comparable performance.
First, we approximate the complicated function $\bar T_{\text{wt}}({\bf A},{\epsilon},\mathcal P) \triangleq \min_{{\bf p} \in \mathcal P} \bar T({\bf A},{\epsilon},{\bf p} )$, which has $N2^K$ terms, with a simpler function, which has $1+\frac{K(K-1)}{2} $ terms.
Specifically, as in Section III.C, we approximate $\bar T({\bf A},{\epsilon},{\bf p} )$ with $\tilde T ({\bf A},{\epsilon},{\bf p} )$.
Recall that $ \tilde T (\mathbf {A},\epsilon,{\bf p})  $ captures the active probabilities of every single device and every two devices.
Analogously, we approximate $\mathcal P$ with
\begin{align}
  \tilde {\mathcal  P} \triangleq  \big\{ (y_{\bf x})_ {{\bf x}\in \mathcal X} \big|\
& \sum\limits_{{\bf x} \in \mathcal X } y_{\bf x} =1, \ \sum\limits_{{\bf x} \in \mathcal X} \underline{p}_{\bf x}x_k
\leq
\sum\limits_{{\bf x} \in \mathcal X} y_{\bf x}x_k
\leq
\sum\limits_{{\bf x} \in \mathcal X}
\overline{p}_{\bf x}x_k,  \ k\in\mathcal K, \nonumber  \\
 & \sum\limits_{{\bf x} \in \mathcal X} \underline{p}_{\bf x}x_kx_{\ell}
\leq
  \sum\limits_{{\bf x} \in \mathcal X} y_{\bf x}x_k x_{\ell}
  \leq
  \sum\limits_{{\bf x} \in \mathcal X}
  \overline{p}_{\bf x}x_k x_{\ell} , \  k,\ell \in\mathcal K ,k<\ell \nonumber \big\}.
  \end{align}
Obviously, $ \tilde {\mathcal  P}  \supseteq \mathcal P $.
Note that contrary to $\mathcal P$, only the upper and lower bounds on the active probabilities of every single device and every two devices remain.
Thus, we approximate $\bar T_{\text{wt}}({\bf A},{\epsilon},\mathcal P) $
with $\tilde T_{\text{wt}}({\bf A},{\epsilon},\tilde {\mathcal  P}  ) \triangleq
\min\nolimits_{{\bf p} \in  \tilde {\mathcal  P} } \tilde T( {\bf A},{\epsilon}, \bf p )$ whose analytical form is given in the following lemma.
\begin{lemma}[Approximate Worst-case Average Throughput] \label{Thm_SubOptLB}
For all $({\bf A},\epsilon)$ satisfying~(\ref{C1}), (\ref{C12}),~(\ref{C2}) and~(\ref{C3}),
\begin{align}
\tilde T_{\text{wt}}({\bf A},{\epsilon}, \tilde {\mathcal  P} )    =
  \epsilon  \sum\limits_{k\in \mathcal {K}} \sum\limits_{{\bf x} \in \mathcal X }
  {\underline p}_{\bf x}  x_k
-\epsilon^2 \sum\limits_{n\in \mathcal {N}} \sum\limits_{k\in \mathcal {K }}a_{k,n} \sum\limits_{\ell \in {\mathcal K}: \ell > k}a_{\ell,n}
\sum\limits_{{\bf x} \in \mathcal X }{\overline p}_{\bf x} x_kx_{\ell} .
\label{ApproxImp}
\end{align}
\end{lemma}
\begin{IEEEproof}
 Please refer to Appendix H.
\end{IEEEproof}

Next, we consider the following approximate problem of Problem~\ref{Prob:Imperfectcase}.
\begin{Prob}[Approximate Worst-case Average Throughput Maximization]
\label{Prob:ImperfectcaseLB}
   \begin{align}
   \max\limits_{\bf A, \epsilon}  &\quad \tilde T_{\text{wt}}({\bf A},{\epsilon},\tilde {\mathcal  P} )  \nonumber \\ 
\mathrm{s.t.}  &\quad \text{(\ref{C1})}, \ \text{(\ref{C12})}, \ \text{(\ref{C2})}, \ \text{(\ref{C3})}, \nonumber
 \end{align}
\end{Prob}
where $ \tilde T_{\text{wt}}({\bf A},{\epsilon}, \tilde {\mathcal  P} )$ is given by (\ref{ApproxImp}).

The numbers of variables and constraints of Problem \ref{Prob:ImperfectcaseLB} are $KN+1$ and $2+ K(N+1)$, respectively, which are much smaller than those of Problem \ref{Prob:ImperfectcaseEQP}.
Note that $\sum\nolimits_{{\bf x} \in \mathcal X }\underline p_{\bf x} x_k$ and $\sum\nolimits_{{\bf x} \in \mathcal X }\overline p_{\bf x}x_{\ell} x_k$ ($k<\ell$) represent a lower bound on the active probability of every single device and an upper bound on the active probability of every two devices, respectively, and can be computed in advance.
Obviously, Problem \ref{Prob:ImperfectcaseLB} shares the same form as Problem \ref{Prob:Perfectcase_LB}.
Thus, we can use a low-complexity iterative algorithm, similar to Algorithm~\ref{Alg_PLBBCD}, to obtain a stationary point of Problem~\ref{Prob:ImperfectcaseLB}.
The details are omitted due to the page limitation.

\section{Stochastic Optimization for Unknown Joint Device Activity Distribution}

In this section, we consider the sample average throughput maximization in the case of unknown joint device activity distribution.
We first formulate the sample average throughput maximization problem, which
 is a challenging nonconvex problem.
Then, we develop an iterative algorithm to obtain a stationary point.

\subsection{Problem Formulation}

In the case of unknown joint device activity distribution, we optimize the
 preamble selection distributions $\bf A$ and access barring factor $\epsilon$ to
 maximize the sample average throughput $ \bar T_{\text{st}}({\bf A},{\epsilon}, \bf p )$ in (\ref{T_st}) subject to the constraints on $({\bf A},\epsilon)$ in (\ref{C1}), (\ref{C12}), (\ref{C2}), and (\ref{C3}).

\begin{Prob}[Sample Average Throughput Maximization]
\label{Prob:Unknowncase}
\begin{align}
   \max _{\bf{A},\epsilon}& \quad  \bar T_{\text{st}}({\bf A},{\epsilon}) \nonumber\\
  \mathrm{s.t.}& \quad   \text{(\ref{C1})},\ \text{(\ref{C12})},  \ \text{(\ref{C2})}, \  \text{(\ref{C3})}.\nonumber
  \end{align}
  \end{Prob}

The objective function $\bar T_{\text{st}}({\bf A},{\epsilon})$ is nonconcave
 in $({\bf A},\epsilon)$, and the constraints in (\ref{C1}), (\ref{C12}),~(\ref{C2}), and~(\ref{C3}) are linear.
Thus, Problem~\ref{Prob:Unknowncase} is nonconvex with a convex feasible set.
	Note that the objective function $\bar T_{\text{st}}({\bf A},{\epsilon})$ has $NKI$ terms, and the number of samples $I$ is usually quite large in practice.
	Therefore, directly tackling Problem~\ref{Prob:Unknowncase} is not computationally efficient.
	To reduce the computation time, we solve its equivalent stochastic version (which is given in Appendix I) using a stochastic iterative algorithm.

\subsection{Stationary Point}
Based on mini-batch stochastic parallel SCA~\cite{koppel2018parallel}, we propose a stochastic algorithm to obtain a stationary point of Problem~\ref{Prob:Unknowncase}.
The main idea is to solve a set of parallelly refined convex problems, each of which is obtained by approximating $\bar T_{\text{st}}({\bf A},{\epsilon})$ with a convex function based on its structure and samples in a uniformly randomly selected mini-batch.
We partition $\mathcal I$ into $M$
disjoint subsets, $\mathcal I_{m}, m\in \mathcal M$, each of size $\frac{I}{M}$ (assuming $I$ is divisible by $M$), where $\mathcal M  \triangleq \{1,2,...,M\}$.
We divide the variables $\left({\bf A}, \epsilon\right)$
into $K+1$ blocks, i.e., ${\bf a}_{k}, \ k \in \mathcal K $ and $ \epsilon $.
 This algorithm updates all $K+1$ blocks in each iteration separately in a parallel manner by maximizing $K+1$ approximate functions of
$\bar T_{\text{st}}({\bf A},{\epsilon})$.

Specifically, at iteration $t$, a mini-batch denoted by $\mathcal I_{\xi^{(t)}}$ is selected, where ${\xi^{(t)}} $ follows the uniform distribution over $\mathcal M$.
Let ${\bf a}_k^{(t-1)}$
and $\epsilon^{(t-1)}$ represent the preamble selection distribution of device $k$ and the access barring factor obtained at iteration $t-1$.
Denote ${\bf A}^{(t-1)}\triangleq  ( {\bf a}_k^{(t-1)})_{k\in \mathcal K}$.
For ease of illustration, in the following, we also write
	$(\mathbf {A},\epsilon)$ as $( {\bf a}_{k},{\bf a}_{-k}, \epsilon)$, where ${\bf a}_{-k}\triangleq({\bf a}_{\ell})_{\ell\in {\mathcal K}, \ell\neq k} $.
We choose
  \begin{align}
  \tilde  T_{\text{st},k}^{(t)} \left({\bf a}_k,  {\bf a}_{-k}^{(t-1)}, \epsilon^{(t-1)} \right )
  \triangleq &\rho^{(t)} \frac{M}{I} \sum\limits_{i \in \mathcal I_{{\xi}^{(t)}} }T \left({\bf A}^{(t-1)},{\epsilon}^{(t-1)},{\bf x}_i \right ) + \sum\limits_{n \in \mathcal N} c_{k,n}^{(t)}
  \left(a_{k,n} - a_{k,n}^{(t-1)}\right)    \nonumber
  \end{align}
as an approximate function of $\bar T_{\text{st}}({\bf A},{\epsilon})$ for updating ${\bf a}_k$ at iteration $t$.
Here,
$\rho^{(t)}$ is a positive diminishing stepsize satisfying
\begin{align}
\rho^{(t)}>0, \ \lim\limits_{t\rightarrow\infty} \rho^{(t)} = 0,
\ \sum\limits_{t=1}^{\infty}\rho^{(t)} = \infty,
\ \sum\limits_{t=1}^{\infty}(\rho^{(t)})^2 < \infty, \nonumber
\end{align}
 and $c_{k,n}^{(t)}$ is given by
  \begin{align}
  	c_{k,n}^{(t)}  =
  	 \left(1-\rho^{(t)} \right)c_{k,n}^{(t-1)}
  	\!	+ \!\rho^{(t)}
  		\frac{M}{I}\epsilon^{(t-1)}\sum\limits_{i \in \mathcal I_{{\xi}^{(t)}}  }
	  x_{i,k} g_{k,n}( {\bf A}^{(t-1)}, \epsilon^{(t-1)},{\bf x}_i)
 	,   \nonumber
  \end{align}
where $ 	c_{k,n}^{(0)} = 0 $, $  k \in \mathcal K, n\in \mathcal N$, $x_{i,k} $ represents the $k$-th element of $\mathbf x_i$, and $g_{k,n}( \mathbf A, \epsilon,{\bf x}) $ is given by (\ref{g_{k,n}}).
 We choose
\begin{align}
  \tilde  T_{\text{st},0}^{(t)}\left( \epsilon,{\bf A}^{(t-1)}, \epsilon^{(t-1)}  \right )
  \triangleq
         c_{0}^{(t)} \epsilon  -
 \tau \left( \epsilon -\epsilon^{(t-1)} \right)^2   \nonumber
  \end{align}
as an approximate function of $\bar T_{\text{st}}({\bf A},{\epsilon})$ for updating $\epsilon$ at iteration $t$.
Here, $\tau$ is a positive constant and $ c_{0}^{(t)}$ is given by
	\begin{align}
  c_{0}^{(t)}=
			 (1-\rho^{(t)})c_{0}^{(t-1)}+
			\rho^{(t)}
			\frac{M}{I}\sum\limits_{i \in \mathcal I_{{\xi}^{(t)}}  }
		\sum\limits_{k \in \mathcal K} x_{i,k }\sum\limits_{n \in \mathcal N} a_{k,n}^{(t-1)}	g_{k,n}( {\bf A}^{(t-1)}, \epsilon^{(t-1)},{\bf x}_i)  ,  \nonumber
	\end{align}
where $c_{0}^{(0)} = 0$.

We first solve the following problems for $K+1$ blocks, in a parallel manner.
Given $\left({\bf A}^{(t-1)},{\epsilon}^{(t-1)} \right)$ and $\mathcal I_{\xi^{(t)}}$,
the optimization problem with respect to ${\bf a}_{k}$ is given by
\begin{align}
  \hat{\bf a}_{k}^{(t)}\triangleq \mathop{\arg\max}\limits_{{\bf a}_k  }&  \quad     \tilde  T_{\text{st},k} ^{(t)}   \left({\bf a}_k,  {\bf a}_{-k}^{(t-1)}, \epsilon^{(t-1)}  \right ) , \quad   k \in \mathcal K \label{Prob:Unknowncase_RAO}  \\
    \mathrm{s.t.}&  \quad  \text{(\ref{C2})}, \ \text{(\ref{C3})},\nonumber
 \end{align}
and the optimization problem with respect to $\epsilon$ is given by
\begin{align}
   \hat \epsilon^{(t)}\triangleq \mathop{\arg\max}\limits_{\epsilon}& \quad
  \tilde  T_{\text{st},0}^{(t)}
  \left( \epsilon  ,{\bf A}^{(t-1)}, \epsilon^{(t-1)}  \right )
      \label{Prob:Unknowncase_ACB}  \\
  \mathrm{s.t.} & \quad \      \text{(\ref{C1})}, \ \text{(\ref{C12})}.\nonumber
 \end{align}
Each problem in (\ref{Prob:Unknowncase_RAO}) is an LP with $N$
 variables and $N+1$ constraints.
The problem in (\ref{Prob:Unknowncase_ACB}) is a QP
with a single variable and two constraints.
Based on the structural properties of the optimization
 problems in (\ref{Prob:Unknowncase_RAO}) and (\ref{Prob:Unknowncase_ACB}), we can
 obtain their optimal points.

\begin{theorem}[Optimal Points of Problems
in (\ref{Prob:Unknowncase_RAO}) and (\ref{Prob:Unknowncase_ACB})]\label{Thm_OptimmalSolutionSPSCA}
A set of optimal points of each optimization problem in (\ref{Prob:Unknowncase_RAO}) is given by
\begin{align}
\left\{ {\bf e}_{m} :     m \in  \mathop{\arg\max}_{n\in\mathcal N}
   \quad
    c_{k,n}^{(t)} \right\}     , \ k\in\mathcal K,  \label{UnknownUpdate_a}
\end{align}
and the optimal point of the optimization problem in (\ref{Prob:Unknowncase_ACB}) is given by
\begin{align}
{\min\left(  \max\left(  \epsilon^{(t-1)}+\frac{c_0^{(t)}}{{2\tau}}  ,0 \right),1 \right)}.\label{UnknownUpdate_epsilon}
\end{align}

\begin{algorithm}[t]
	\caption{Obtaining A Stationary Point of Problem \ref{Prob:Unknowncase}}
	\label{Alg_SPSCA}
	\begin{multicols}{2}
		\begin{algorithmic}[1]
			\small{ \STATE \textbf{initialization:}
				for $k \in \mathcal K$, set ${\bf a}_k:= {\bf e}_{n_k}$, where $n_k$ is uniformly chosen from $\mathcal N$ at random, set $\epsilon:=1$, set $t:=0$, and choose $\mu>0$.
				\STATE \textbf{repeat}
				\STATE $t:=t +1$.
				\STATE  \textbf{for $k \in \mathcal K$ do}
				\STATE \quad  $\hat {\bf a}_k $ is randomly chosen from the set in (\ref{UnknownUpdate_a}).
				\STATE \quad Update ${ {\bf a}}_{k}^{(t)} $ according to (\ref{UpdatePSCA_k}).
				\STATE \textbf{end for}
				\STATE \quad Calculate $ \hat\epsilon^{(t)} $ according to (\ref{UnknownUpdate_epsilon}).
				\STATE \quad Update $ \epsilon^{(t)} $ according to (\ref{UpdatePSCA_0}).
				\STATE \textbf{until} $ \| {\bf A}^{(t)} -{\bf A}^{(t-1)}    \|_{\mathbb F} \leq \mu $.
				\normalsize}
		\end{algorithmic}
	\end{multicols}
\end{algorithm}

\end{theorem}

\begin{IEEEproof}
Theorem \ref{Thm_OptimmalSolutionSPSCA} can be proved in a similar way to Theorem \ref{Thm_OptimmalSolutionBCD}. 
 \end{IEEEproof}

The optimal point in (\ref{UnknownUpdate_a}) indicates that each device $k $ selects the preamble corresponding to the maximum approximate average throughput (increase rate of the average throughput) conditioned on selected by device $k$ for given $ (\mathbf a_{-k}^{(t-1)}, \epsilon^{(t-1)})$.
For all $k \in \mathcal K$ and $n \in \mathcal N$, the computational complexity for
calculating $c_{k,n}^{(t)} $ is $\mathcal O (  K^2)$.
For all $k \in \mathcal K$, the computational complexity of finding the largest one among $c_{k,n}^{(t)}$, $n\in \mathcal N$ is $\mathcal O (N)$.
Thus, the overall computational complexity for determining the sets in (\ref{UnknownUpdate_a}) is
$\mathcal O ( NK^3  )  + \mathcal O ( NK)  =\mathcal O (NK^3) $.
The computational complexity for calculating $c_0^{(t)}$ is $  \mathcal O( NK^3)$, which is also the computational complexity for obtaining the optimal point in (\ref{UnknownUpdate_epsilon}).

Then, we update the preamble selection distributions $\bf A$ and access barring factor $\epsilon$ by
\begin{subequations}\label{UpdateRobust}
\begin{align}
&{\bf a}_{k}^{(t)} = (1-\omega^{(t)} ) {\bf a}_{k}^{(t-1)} + \hat{{\bf a}}_{k}^{(t)},\  k\in\mathcal K ,  \label{UpdatePSCA_k}\\
&   \epsilon^{(t)} =   (1-\omega^{(t)} ) \epsilon^{(t-1)} + \hat \epsilon^{(t)},            \label{UpdatePSCA_0}
\end{align}
\end{subequations}
where $\omega^{(t)}$ is a positive diminishing stepsize satisfying
\begin{align}
\omega^{(t)}>0, \ \lim\limits_{t\rightarrow\infty} \omega^{(t)} = 0,
\ \sum\limits_{t=1}^{\infty}\omega^{(t)} = \infty,
\ \sum\limits_{t=1}^{\infty}(\omega^{(t)})^2 < \infty,\ \lim\limits_{t\rightarrow\infty} \frac{\omega^{(t)}}{\rho^{(t)} } = 0.    \nonumber
\end{align}

Finally, the details of the proposed stochastic parallel iterative algorithm are summarized in Algorithm~\ref{Alg_SPSCA}.
Based on \cite{Yang2016parallel}, we can show the following result.
\begin{theorem}[Convergence of Algorithm \ref{Alg_SPSCA}] \label{Thm_SPSCA}
If $\mathop{\arg\max}_{n\in\mathcal N}c_{k,n}^{(t)}$ is a singleton for all $k\in\mathcal K$ and all $t \geq 1$, then every limit point of $\{\left(
{\bf A}^{(t)},\epsilon^{(t)}\right)\}$ generated by Algorithm \ref{Alg_SPSCA} is a stationary point of Problem \ref{Prob:Unknowncase}
almost surely.
\end{theorem}

\begin{IEEEproof}
 Please refer to Appendix I.
 \end{IEEEproof}

\section{Numerical Results}
In this section, we evaluate the performance of the proposed
solutions\footnote{Algorithm~\ref{Alg_PLBBCD} and Algorithm~\ref{Alg_SPSCA} are computationally efficient and can be easily implemented in practical systems with large $K$.
	Algorithm~\ref{Alg_PBCD} and Algorithm~\ref{Alg_ROpt}
	are computationally expensive but can provide essential benchmarks for designing effective and low-complexity methods.} for dependent and independent device activities via numerical results.
In the simulation, we adopt the group device activity model in~\cite{jiang2021ml}.
Specifically, $K$ devices are divided into $G$ groups,
 each of size $\frac{K}{G}$ (assuming $K$ is divisible by $G$),
 the activity states of devices in different groups are independent,
 and the activity states of devices in one group are the same.
 Note that when $\frac{K}{G}=1$, the device activities are independent; when $\frac{K}{G}>1$, the device activities are dependent.
Denote $ \mathcal G \triangleq \{1,2,...,G\}$.
 	Let $\mathcal K_g$ denote the set of devices in group $g \in \mathcal G$.
 Let $y_g \in \{0,1\}  $ denote the activity state of group $g$,
 where $y_g=1$ if group $g$ is active, and $y_g=0$ otherwise.
 Let ${\bf y }  \triangleq (y_g)_{g \in \mathcal G}$ denote the activity states of all groups.
The probability that a group is active is $p_a$.
Then, in the case of perfect joint device activity distribution, the joint device activity distribution is given by
	\begin{align}
		p_{\bf x} = \left\{
		\begin{array}{ll}
			{p_a}^{\sum\limits_{g \in \mathcal G}y_g} (1-p_a)^{ G -\sum\limits_{g \in \mathcal G}y_g }      &, \ x_k = y_g, k \in \mathcal K_g , g \in \mathcal G, {\bf y} \in \{0,1\}^G \\
			0  &, \ \text{otherwise}
		\end{array}
		\right. , \  {\bf x} \in \mathcal X.  \label{Sim_p_x}
	\end{align}	
In the case of imperfect joint device activity distribution, the estimated joint device activity distribution is given by
		\begin{align}
		\hat p_{\bf x} = \left\{
		\begin{array}{ll}
			{p_a}^{\sum\limits_{g \in \mathcal G}y_g} (1-p_a)^{ G -\sum\limits_{g \in \mathcal G}y_g }      &, \ x_k = y_g, k \in \mathcal K_g , g \in \mathcal G, {\bf y} \in \{0,1\}^G \\
			0  &, \ \text{otherwise}
		\end{array}
		\right. , \  {\bf x} \in \mathcal X,   \nonumber
	\end{align}
for $K\leq100$, we set $\delta_{\bf x} = \bar\delta{\hat p}_{\bf x}$, ${\bf x} \in \mathcal X$, and for $K>100$, we set
 \begin{align}
 \delta_{\bf x} =
  \left\{\begin{array}{ll}
 \bar\delta {\hat p}_{\bf x}  &, \  x_k = y_g, k \in \mathcal K_g , g \in \mathcal G, \sum\limits_{g \in \mathcal G}y_g \leq 3, {\bf y} \in \{0,1\}^G \\
   0 &, \ \text{otherwise}
   \end{array}
   \right. , \  {\bf x} \in \mathcal X,   \nonumber
 \end{align}
 for some $\bar\delta \in [0,1)$.
 In the case of unknown joint device activity distribution, we consider $10^5$ samples generated according to the joint device activity distribution $p_{\bf x}, { \bf x} \in \mathcal X$ given by (\ref{Sim_p_x}).
 Note that in the three cases, we consider the same underlying joint device activity distribution $\mathbf p$ for a fair comparison.
 For ease of presentation, in the following, $\bar T({\bf A},{\epsilon}, {\bf p} )$, $\bar T_{\text{wt}}({\bf A},{\epsilon},\mathcal P  )$, and $\bar T_{\text{st}}({\bf A},{\epsilon})$'s numerical mean in
the cases of perfect, imperfect, and unknown joint device activity distributions, i.e., case-pp, case-ip, and case-up, respectively, are referred to as throughput, unless otherwise specified.
We evaluate $\bar T_{\text{st}}({\bf A},{\epsilon})$'s numerical mean by averaging over $100$ realizations of $\bar T_{\text{st}}({\bf A},{\epsilon})$, each obtained based on a set of $10^5$ samples.
In case-$t$, the proposed stationary point or KKT point is called Prop-$t$, where $t = $ pp, ip, up.
Furthermore, in case-$t$, the proposed low-complexity solution is called Prop-LowComp-$t$, where $t = $ pp and ip.

We consider three baseline schemes, namely BL-MMPC, BL-MSPC and BL-LTE.
In BL-MMPC and BL-MSPC, $\bf A$ are obtained by the MMPC and MSPC allocation algorithms and the proposed access barring scheme in \cite{Popovski18SPAWC}, respectively.
In BL-LTE, we set $a_{k,n} = \frac{1}{N}, \ k\in\mathcal K, n\in\mathcal N$ according to the standard random access procedure of LTE networks \cite{TR_3gppevolved} and set $\epsilon= \min \left(1,\frac{N}{\bar K}\right)$ according to the optimal access control \cite{Koucheryavy_ISIT2013}, where $\bar K$ denotes the average number of active devices. 
Note that BL-MMPC and BL-MSPC make use of the dependence of the activities of every two devices; BL-LTE does not utilize any information on the dependence of device activities.
In case-$t$, the three baseline schemes are referred to as BL-MMPC-$t$, BL-MSPC-$t$, and BL-LTE-$t$, where $t = $ pp, ip, up.
In case-pp, ${(\bf A},\epsilon)$ in BL-MMPC-pp and BL-MSPC-pp and $\epsilon$ in BL-LTE-pp rely on $p_{\bf x} $, ${\bf x} \in \mathcal X$.
In case-ip, ${(\bf A},\epsilon)$ in BL-MMPC-ip and BL-MSPC-ip and $\epsilon$ in BL-LTE-ip rely on $\hat p_{\bf x} $, ${\bf x} \in \mathcal X$ without considering potential estimation errors.
In case-up, ${(\bf A},\epsilon)$ in BL-MMPC-up and BL-MSPC-up and $\epsilon$ in BL-LTE-up rely on the empirical joint device activity distribution obtained from the samples.
Considering the tradeoff between the throughput and computational complexity, we run each proposed algorithm five times to obtain each point for all considered parameters.

\begin{figure}[t]
	\begin{center}
		\subfigure[\footnotesize{Throughput versus $K$ at $p_a=0.25$ and $N=15$. }]
		{\resizebox{5cm}{!}{\includegraphics{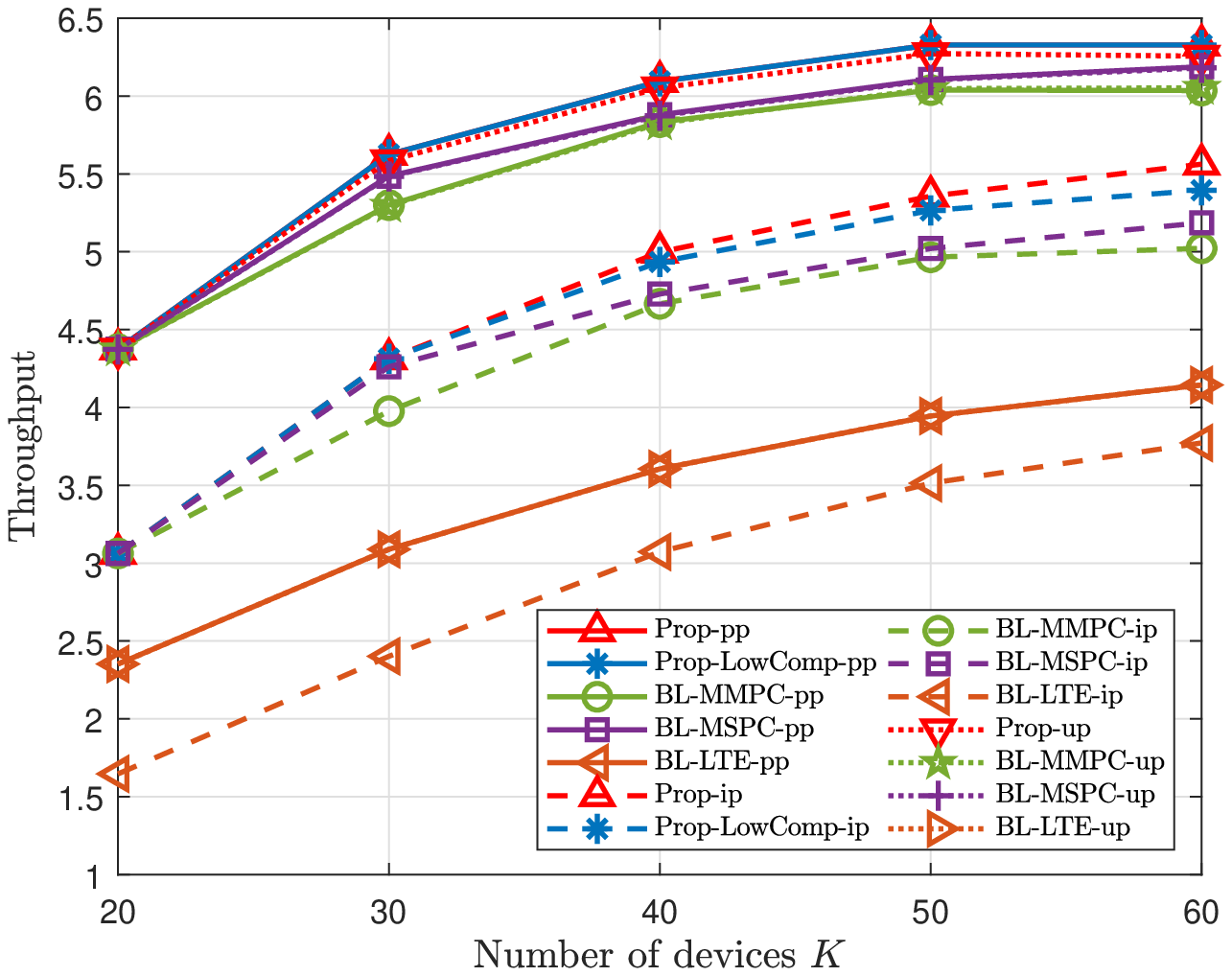}}} \quad
		\subfigure[\footnotesize{Throughput versus $N$ at $p_a=0.25$ and $K=60$.}]
		{\resizebox{5cm}{!}{\includegraphics{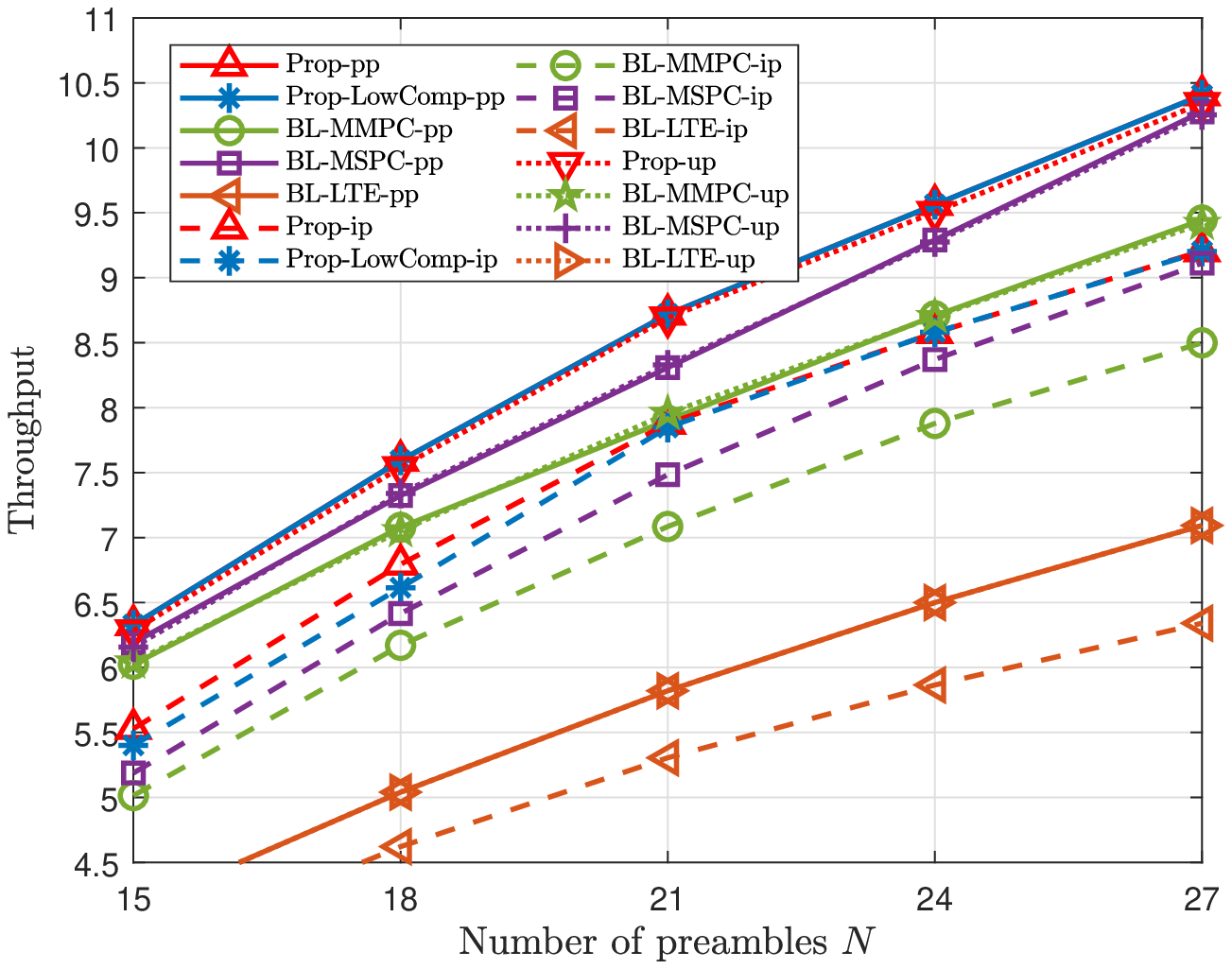}}}
		\quad
		\subfigure[\footnotesize{Throughput versus $p_a$ at $N = 15$ and $K=60$.}]
		{\resizebox{5cm}{!}{\includegraphics{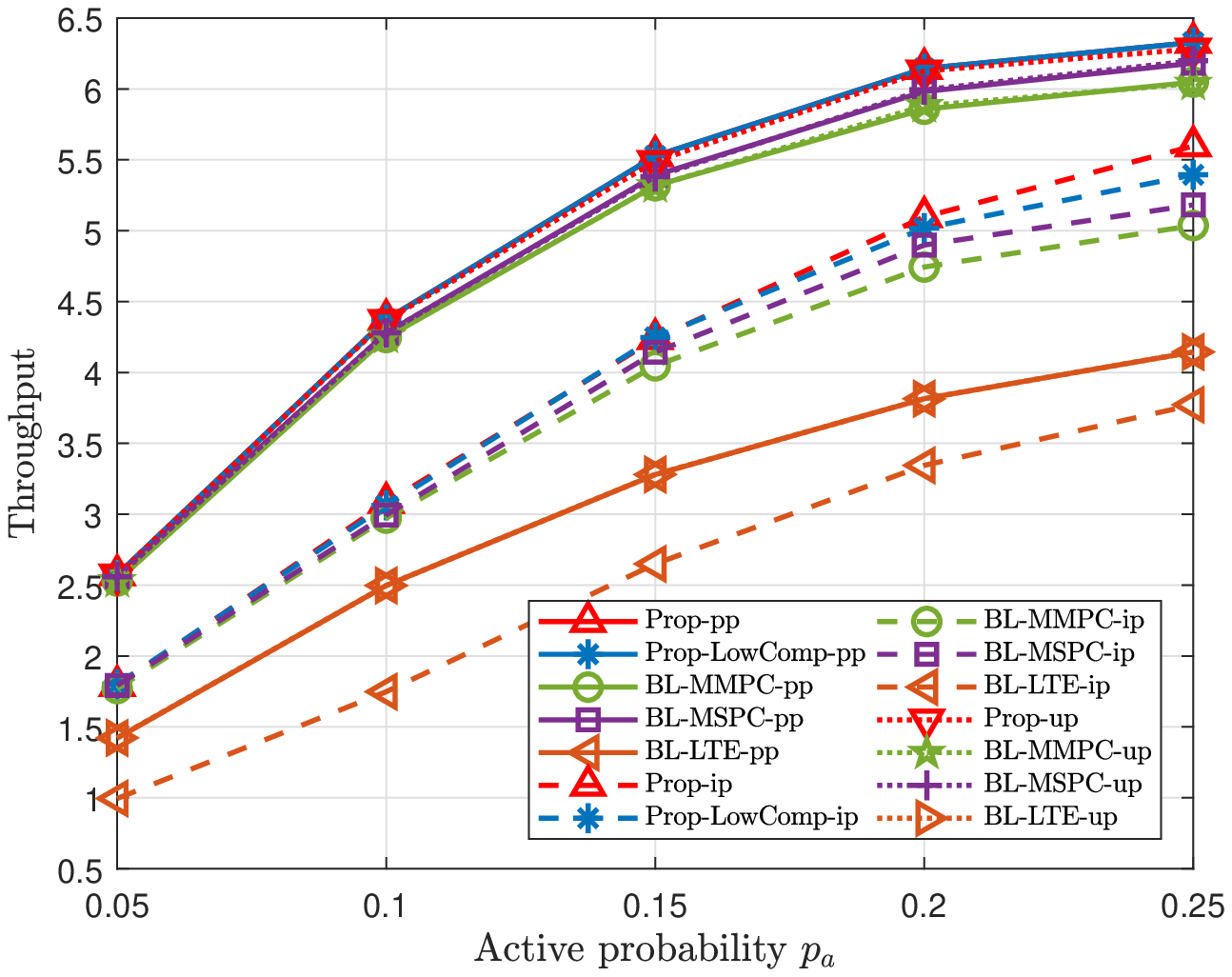}}}
	\end{center}
	\vspace{-3mm}
	\caption{\footnotesize{Throughput versus $K$, $N$, and $p_a$ at $\bar \delta= 0.3$ and $\frac{K}{G} =10$ (dependent device activities).
	}}
	\vspace{-8mm}
	\label{Figure_Throughput_small_K_correlated}
	
\end{figure}

\begin{figure}[t]
	\begin{center}
		\subfigure[\footnotesize{Throughput versus $K$ at $p_a=0.25$ and $N=4$. }]
		{\resizebox{5cm}{!}{\includegraphics{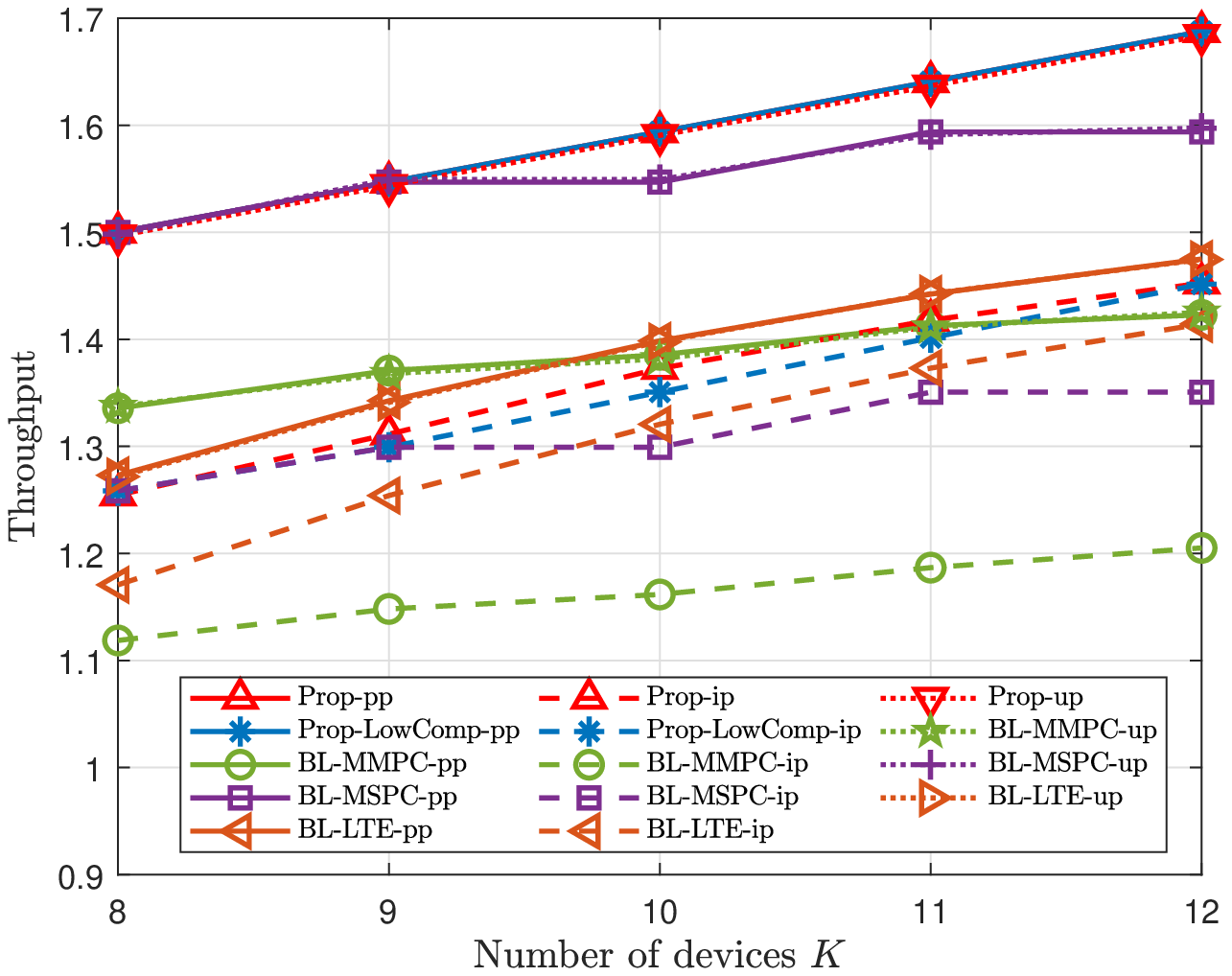}}} \quad
		\subfigure[\footnotesize{Throughput versus $N$ at $p_a=0.25$ and $K=10$.}]
		{\resizebox{5cm}{!}{\includegraphics{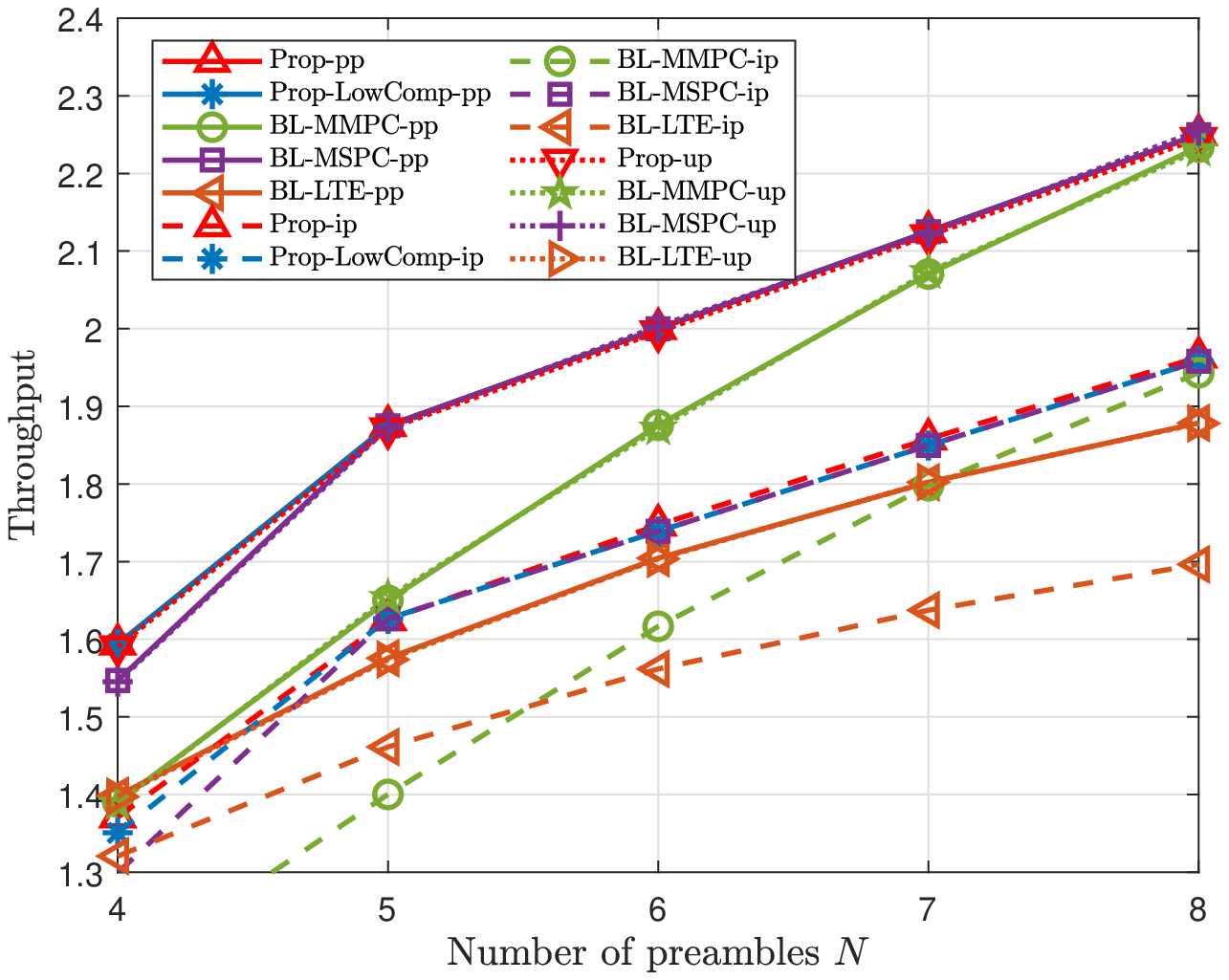}}}
		\quad
		\subfigure[\footnotesize{Throughput versus $p_a$ at $N = 4$ and $K=10$.}]
		{\resizebox{5cm}{!}{\includegraphics{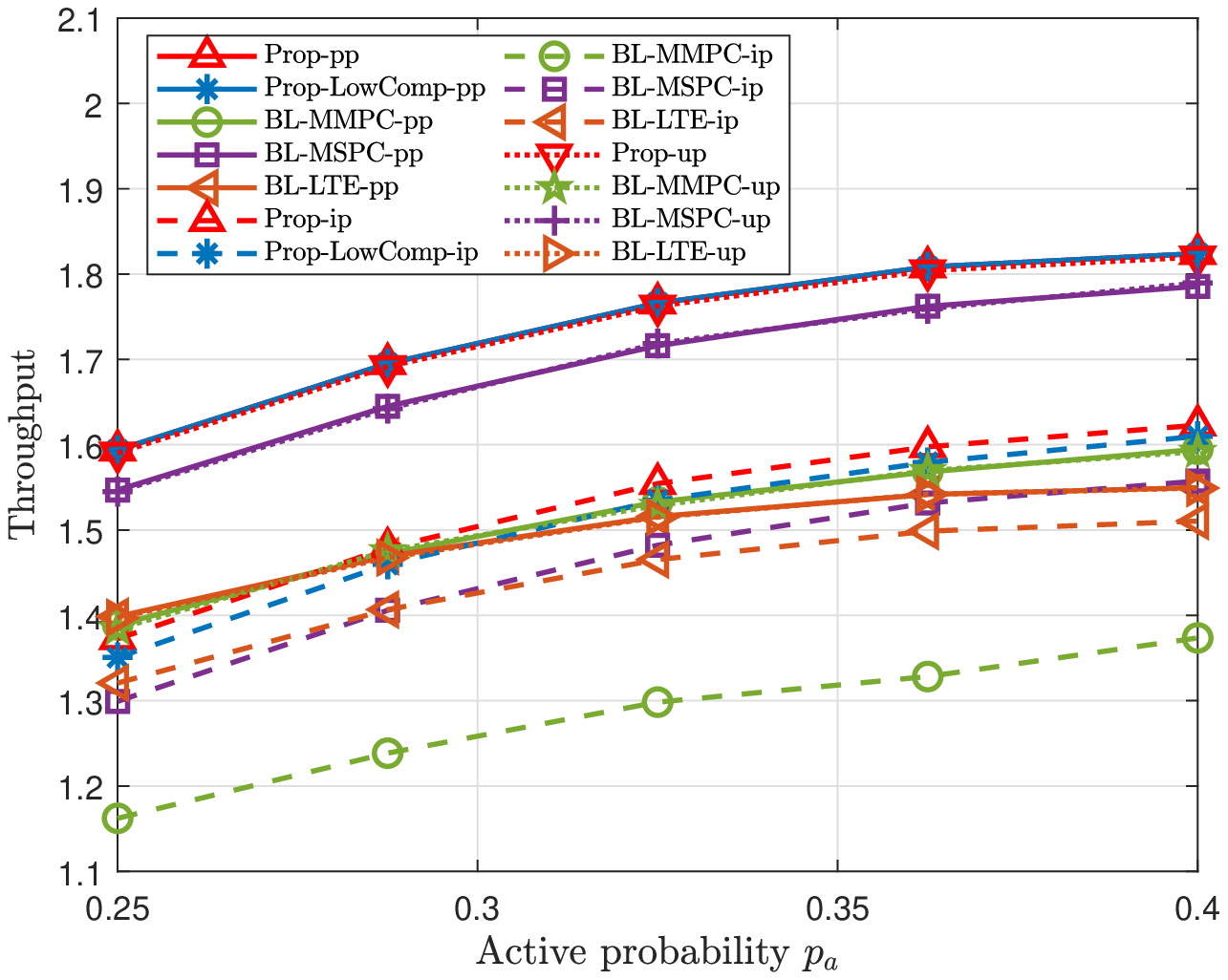}}} 
	\end{center}
	\vspace{-3mm}
	\caption{\footnotesize{Throughput versus $K$, $N$, and $p_a$ at $\bar \delta = 0.3$ and $\frac{K}{G} =1$ (independent device activities).
	}}
	\vspace{-8mm}
	\label{Figure_Throughput_small_K_uncorrelated}
	
\end{figure}

\subsection{Small K and N}
This part compares the throughputs of all proposed solutions and three baseline schemes at small numbers of devices and preambles.
Fig.~\ref{Figure_Throughput_small_K_correlated} and Fig.~\ref{Figure_Throughput_small_K_uncorrelated} illustrate the throughput versus
the number of devices $K$, the number of preambles $N$, and the group active probability $p_a$, for dependent and independent device activities, respectively. 
From Fig.~\ref{Figure_Throughput_small_K_correlated} and Fig.~\ref{Figure_Throughput_small_K_uncorrelated}, we make the following observations.
For each scheme, the curve in case-up is close to that in case-pp, as we evaluate $\bar T_{\text{st}}({\bf A},{\epsilon})$'s numerical mean over a large number of samples;
	the curve in case-pp is above that in case-ip, as illustrated in Section II.
For $t =$ pp, ip, and up, Prop-$t$ outperforms all the baseline schemes in case-$t$, as Prop-$t$ utilizes more statistical information on device activities.
For $t =$ pp and ip, Prop-LowComp-$t$ outperforms all the baseline schemes in case-$t$, as Prop-LowComp-$t$ relies on a more accurate approximation of the throughput.
The fact that the gap between the throughputs of Prop-$t$ and Prop-LowComp-$t$ is small, where $t=$ pp and ip, 
 shows that exploiting statistical information on the activities of every two devices rigorously already achieves a significant gain.
	Furthermore,
from Fig.~\ref{Figure_Throughput_small_K_correlated}~(a), Fig.~\ref{Figure_Throughput_small_K_uncorrelated}~(a) and 
Fig.~\ref{Figure_Throughput_small_K_correlated}~(c), Fig.~\ref{Figure_Throughput_small_K_uncorrelated}~(c), we can see that the throughput of each proposed scheme increases with $K$ and $p_a$, respectively, due to the increase of traffic load.
From Fig.~\ref{Figure_Throughput_small_K_correlated}~(b) and Fig.~\ref{Figure_Throughput_small_K_uncorrelated}~(b), we can see that the throughput of each scheme increases with $N$ due to the increase of communications resource.

\begin{figure}[htbp]
	\centering
	\begin{minipage}[t]{0.48\textwidth}
		\centering
		\includegraphics[width=0.8\textwidth]{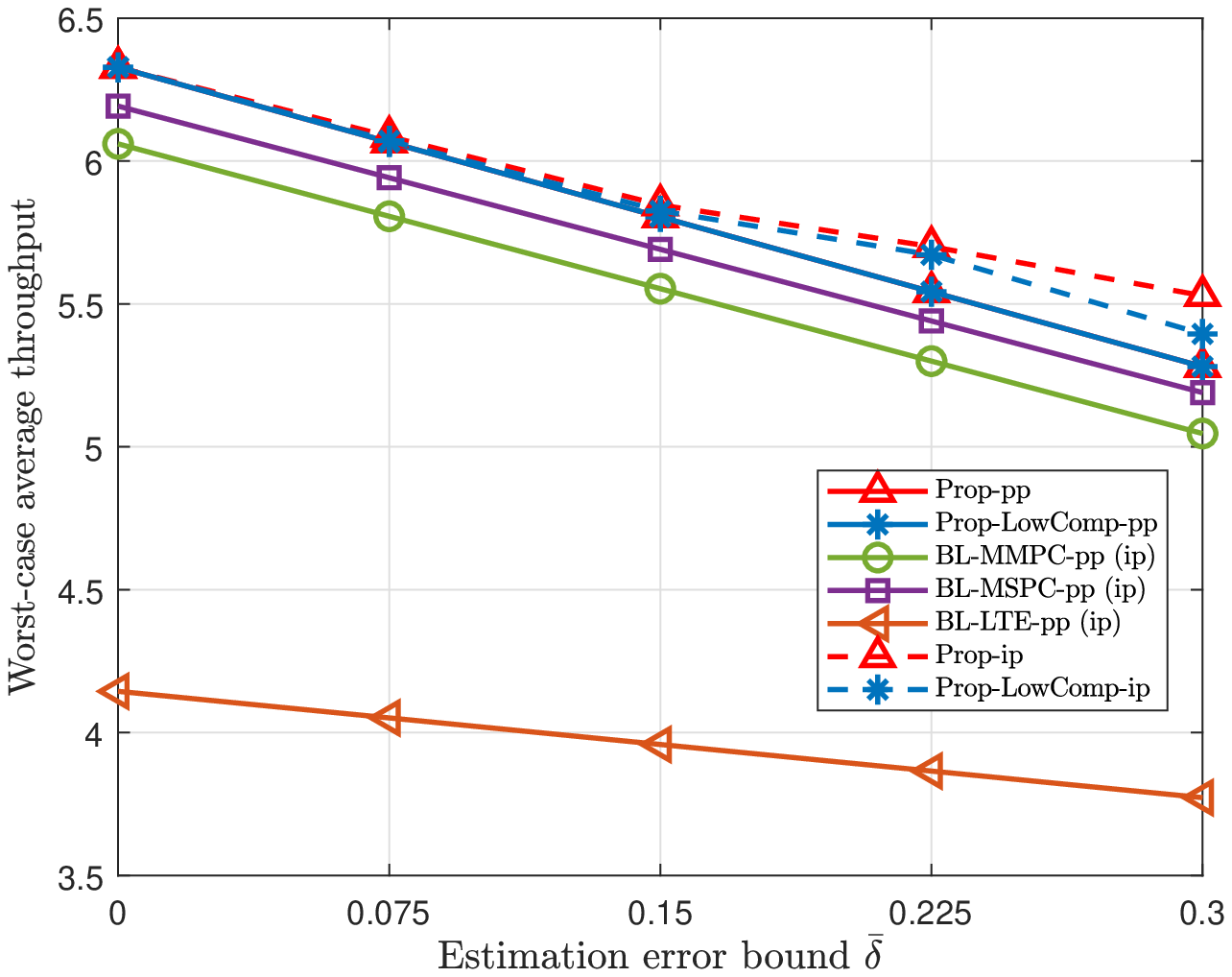}
		\vspace{-3mm}
		\caption{\small{Worst-case average throughput versus $\bar\delta$ at $K=60$, $N=15$, $p_a=0.25$, and $\frac{K}{G} =10$ (dependent device activities).}}\label{Figure_WorstThroughput_small_K_correlated}
	\end{minipage}
     \  
	\begin{minipage}[t]{0.48\textwidth}
		\centering
	\includegraphics[width=0.8\textwidth]{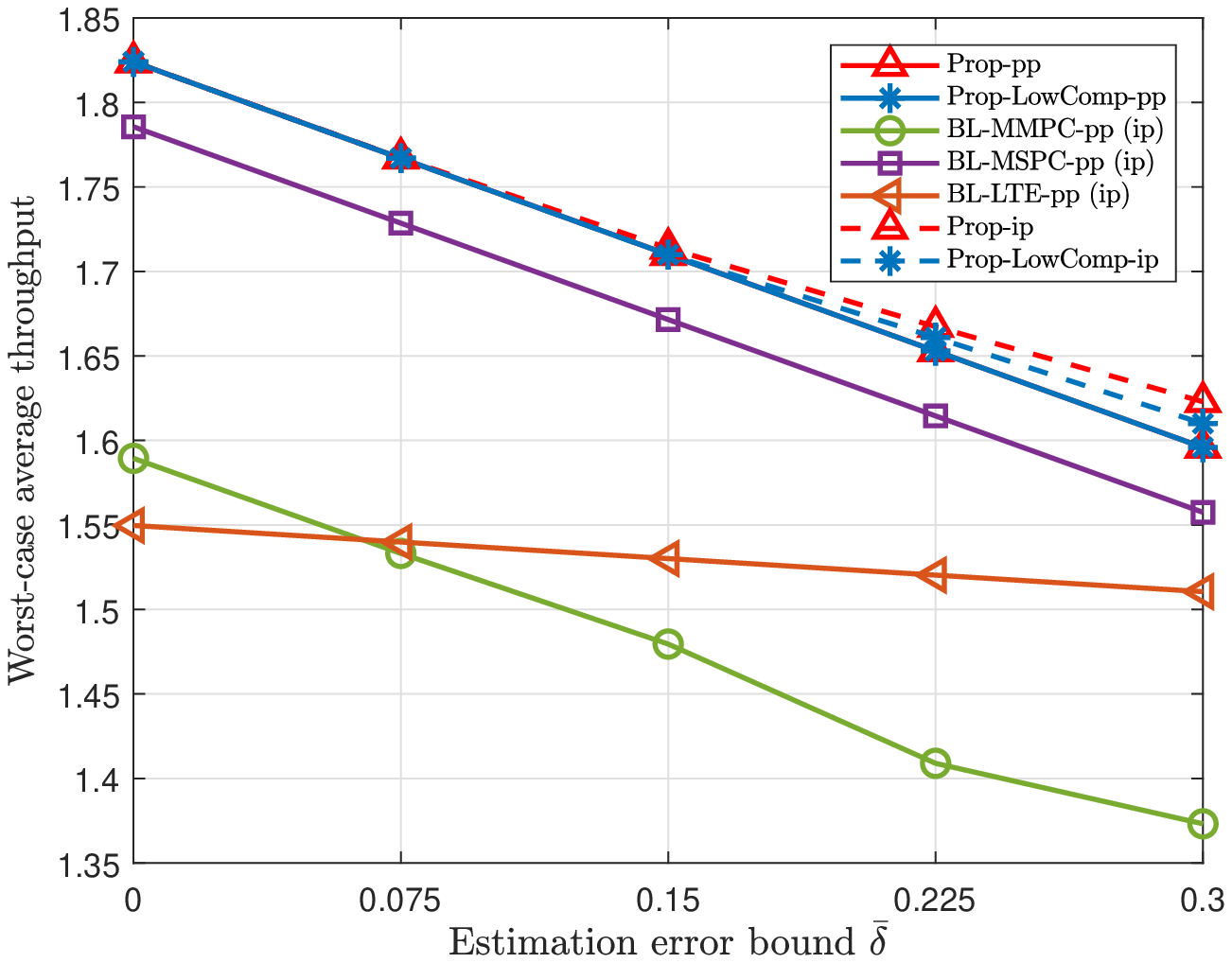}
		\vspace{-3mm}
		\caption{\small{Worst-case average throughput versus $\bar\delta$ at $K=10$, $N=4$, $p_a=0.4$, and $\frac{K}{G} =1$ (independent device activities).}}\label{Figure_WorstThroughput_small_K_uncorrelated}
	\end{minipage}
\end{figure}

Fig.~\ref{Figure_WorstThroughput_small_K_correlated} and Fig.~\ref{Figure_WorstThroughput_small_K_uncorrelated} illustrate the worst-case average throughput versus the estimation error bound $\bar \delta$ for dependent and independent device activities, respectively.
From Fig.~\ref{Figure_WorstThroughput_small_K_correlated} and Fig.~\ref{Figure_WorstThroughput_small_K_uncorrelated}, we can see the worst-case average throughputs of Prop-ip and Prop-LowComp-ip are greater than those of Prop-pp, Prop-LowComp-pp, BL-MMPC-pp, BL-MSPC-pp, and BL-LTE-pp, which reveals the importance of explicitly considering the imperfectness of the estimated joint device activity distribution in this case.
 Furthermore, the gain of Prop-ip over Prop-pp and the gain of Prop-LowComp-ip
 over Prop-LowComp-pp increase with $\bar \delta$, as it is more important to take into account possible device activity estimation errors when $\bar \delta$ is larger.

 \begin{figure}[t]
 	\begin{center}
 		\subfigure[\small{Throughput versus $K$ at $p_a = 0.03$ and $N =50$.}]
 		{\resizebox{5 cm}{!}{\includegraphics{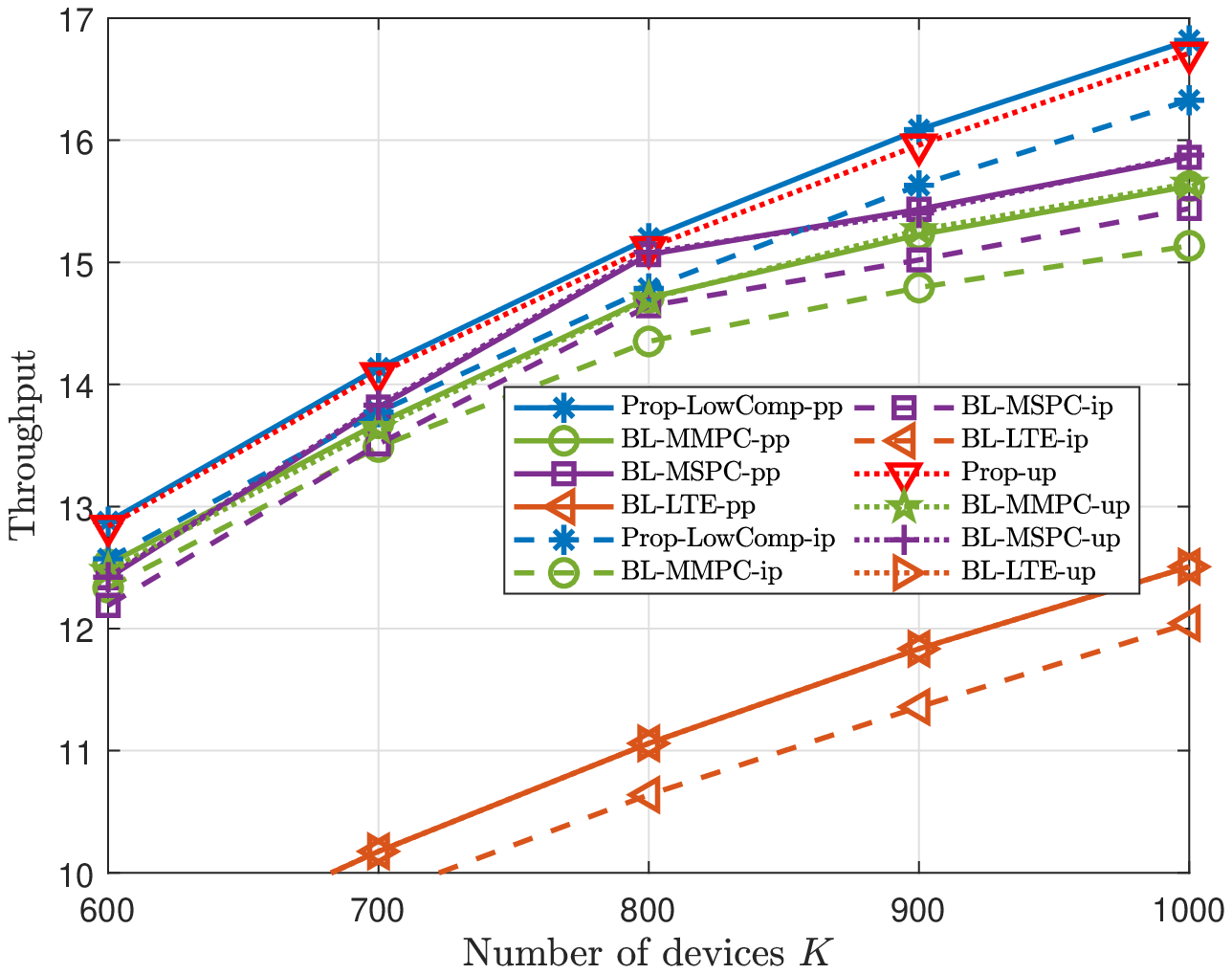}}}\quad
 		\subfigure[\small{Throughput versus $N$ at $p_a =0.03$ and $K = 1000$. }]
 		{\resizebox{5 cm}{!}{\includegraphics{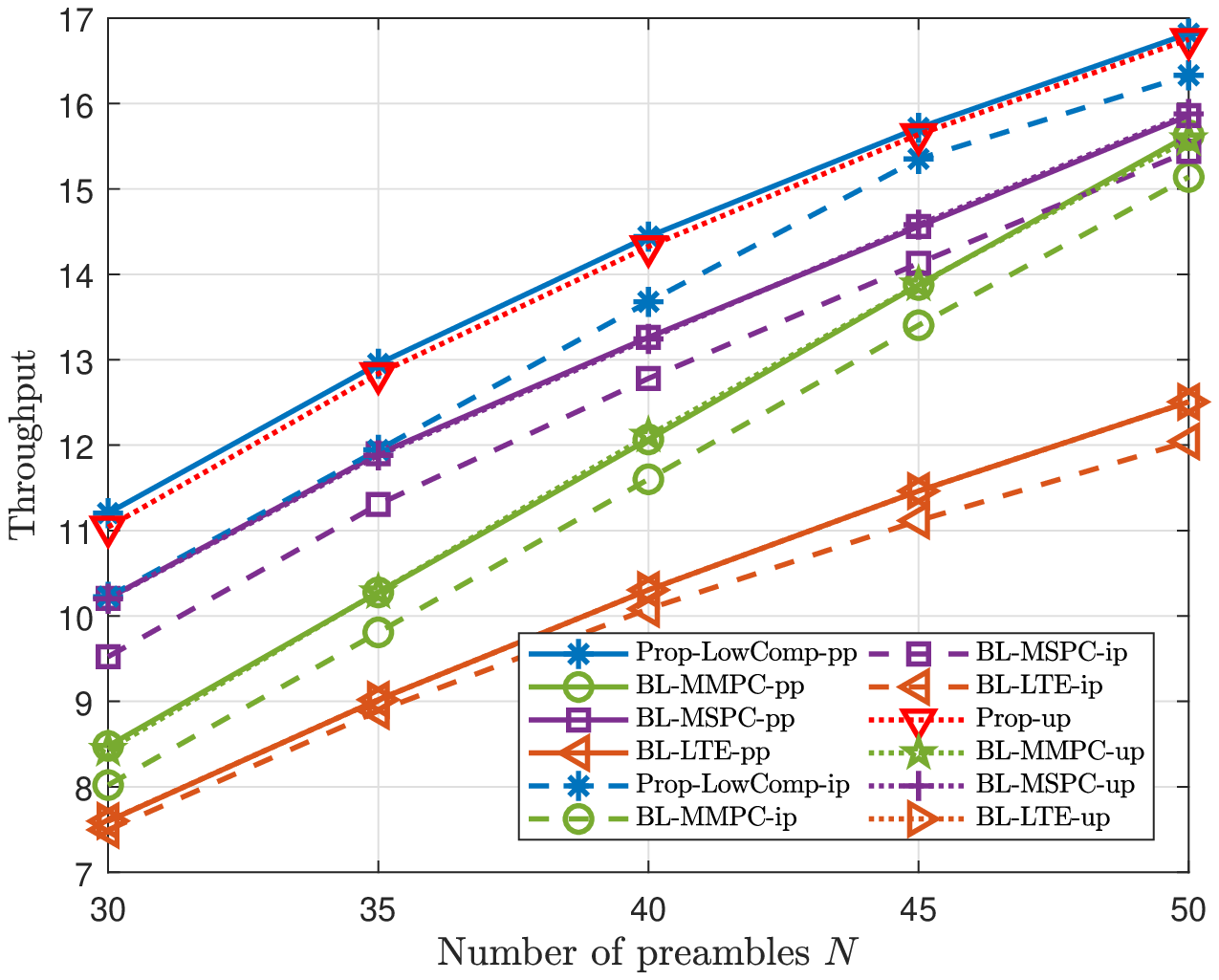}}} \quad
 		\subfigure[\small{Throughput versus $p_a$ at $N = 50$ and $K =1000$.}]
 		{\resizebox{5cm}{!}{\includegraphics{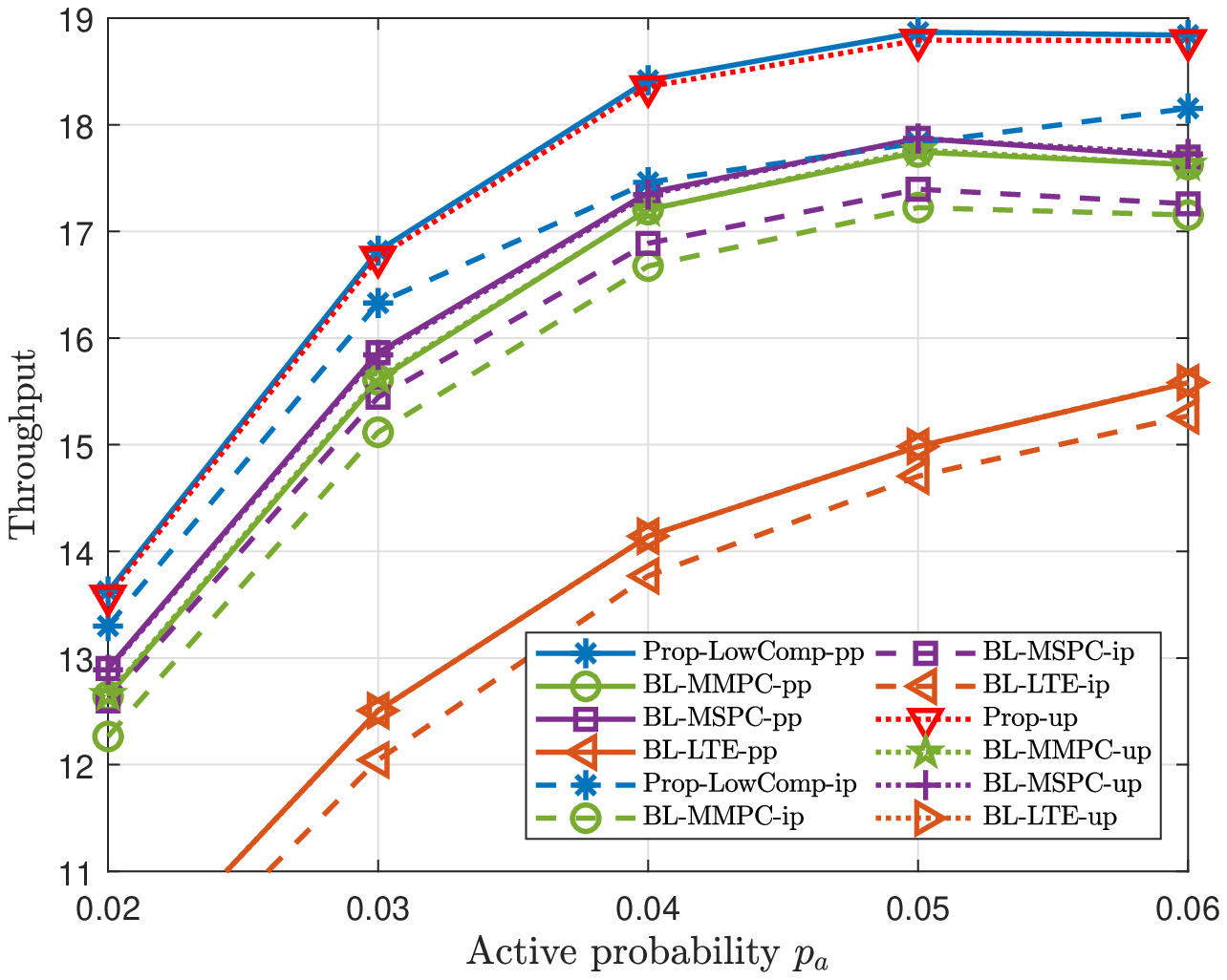}}} 
 	\end{center}
 	\vspace{-5mm}
 	\caption{{Throughput versus $K$, $N$, and $p_a$ at $\bar \delta = 0.3$ and $\frac{K}{G} =20$ (dependent device activities).
 	}}
 	\vspace{-8mm}
 	\label{Figure:TotalUtility_correlated}
 \end{figure}

\begin{figure}[t]
	\begin{center}
		\subfigure[\small{Throughput versus $K$ at $p_a = 0.03$ and $N =50$.}]
		{\resizebox{5 cm}{!}{\includegraphics{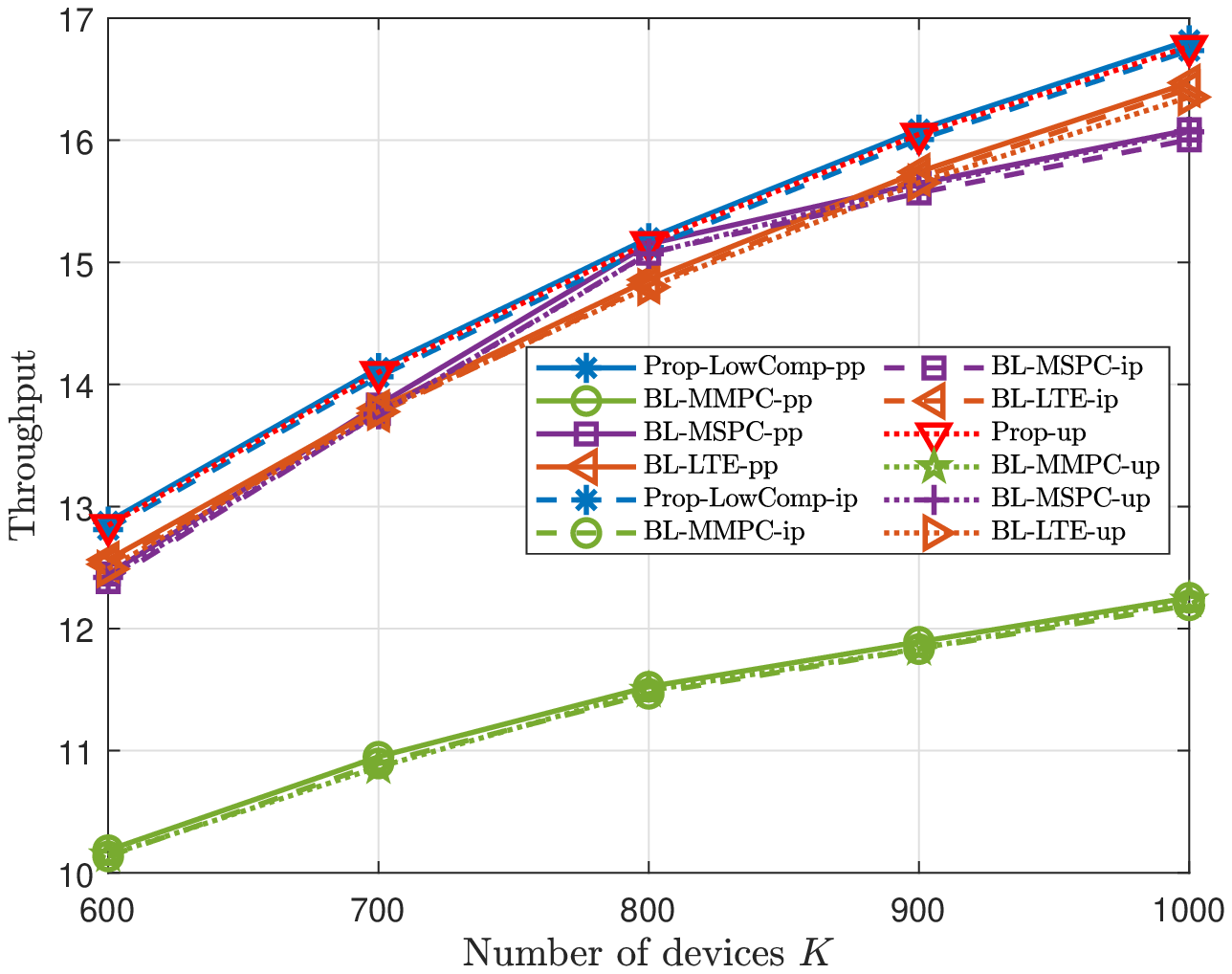}}}\quad
		\subfigure[\small{Throughput versus $N$ at $p_a =0.03$ and $K = 1000$. }]
		{\resizebox{5 cm}{!}{\includegraphics{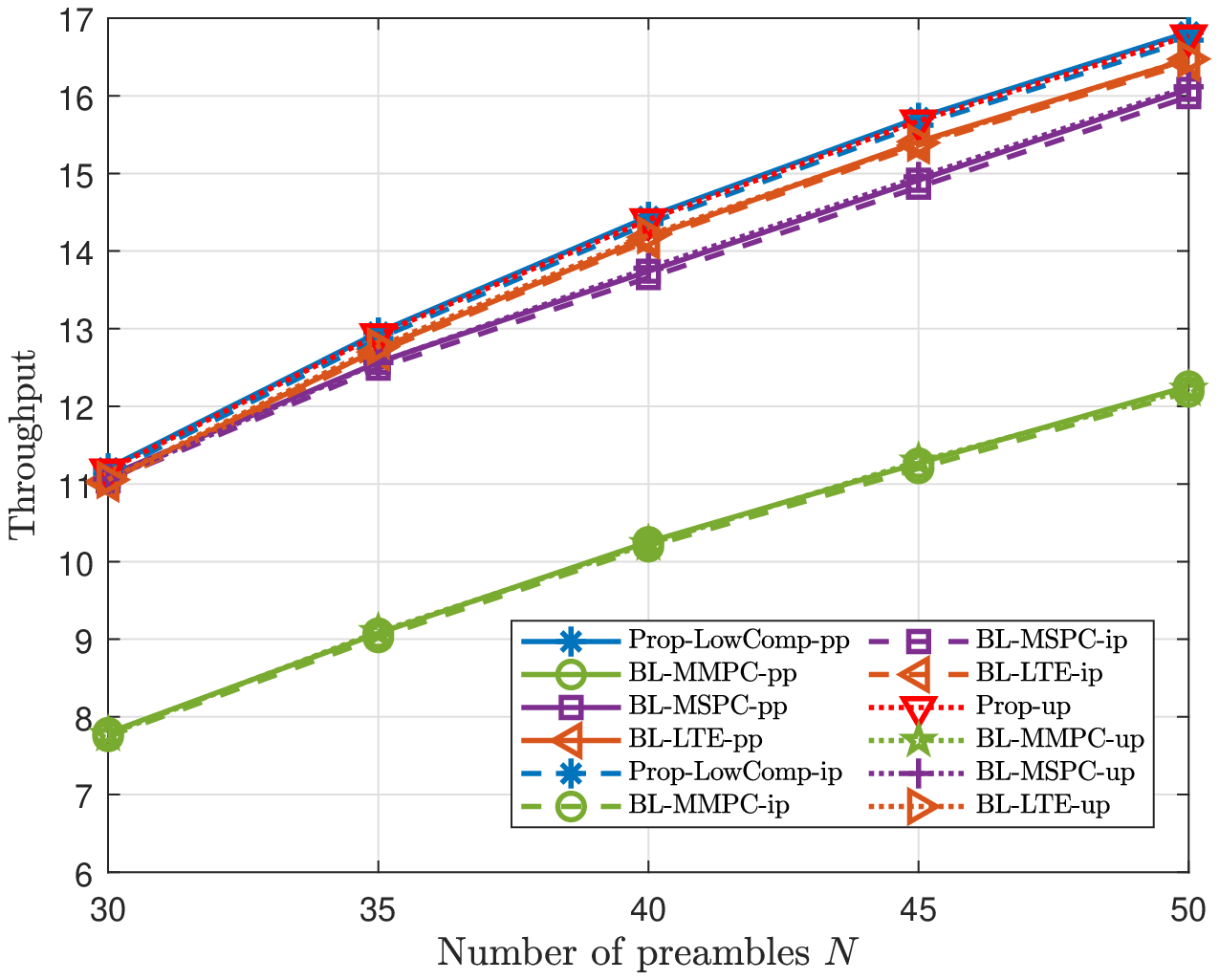}}} \quad
		\subfigure[\small{Throughput versus $p_a$ at $N = 50$ and $K =1000$.}]
		{\resizebox{5cm}{!}{\includegraphics{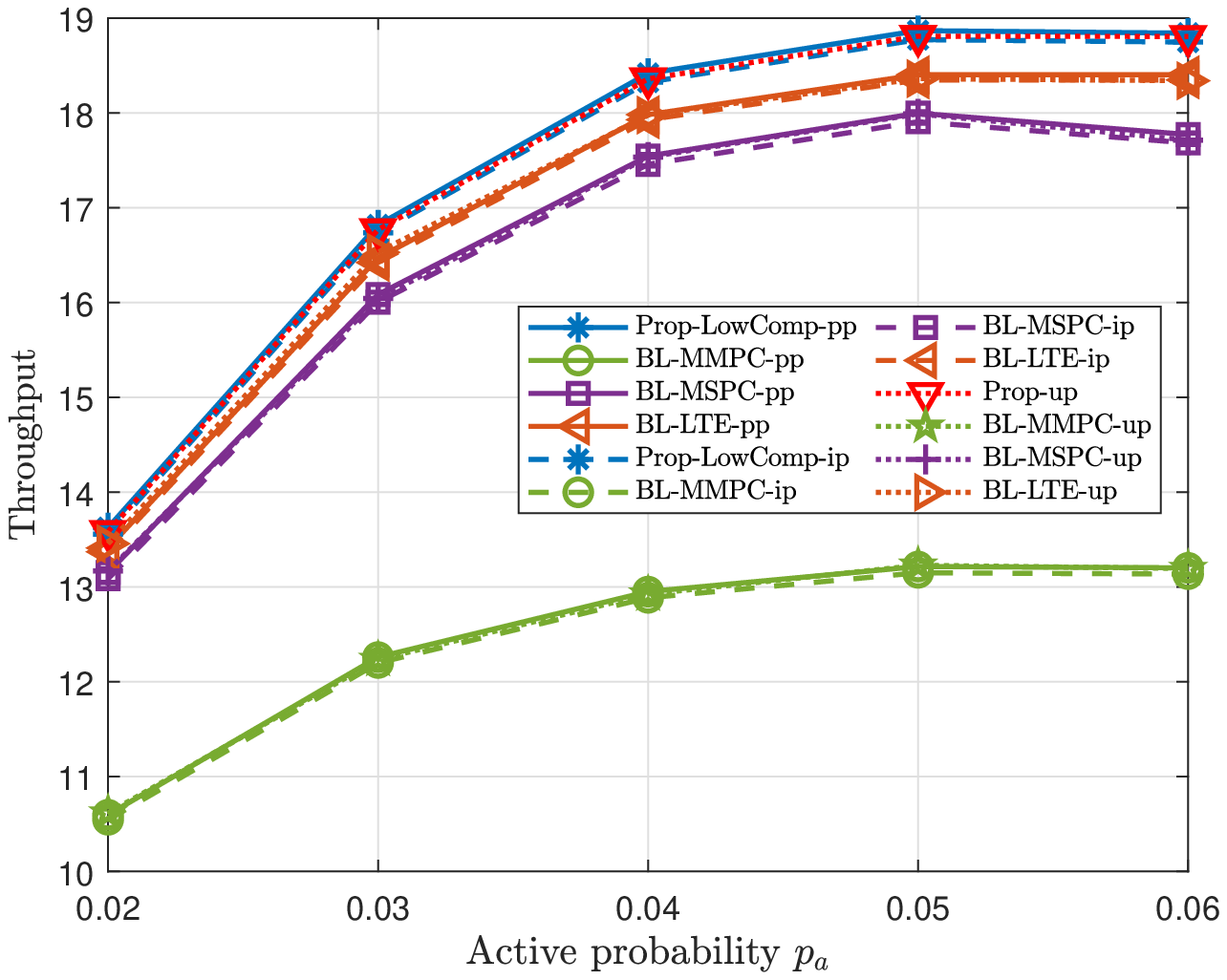}}}
	\end{center}
	\vspace{-5mm}
	\caption{{Throughput versus $K$, $N$, and $p_a$ at $\bar\delta = 0.3$ and $\frac{K}{G} =1$ (independent device activities).
	}}
	\vspace{-10mm}
	\label{Figure:TotalUtility_uncorrelated}
\end{figure}

\subsection{Large K and N}
This part compares the throughputs of Prop-LowComp-pp, Prop-LowComp-ip, Prop-up, and three baseline schemes, at large numbers of devices and preambles.
Fig.~\ref{Figure:TotalUtility_correlated} and Fig.~\ref{Figure:TotalUtility_uncorrelated} illustrate the throughput versus
	the number of devices $K$, the number of preambles $N$, and the group active probability $p_a$, for dependent and independent device activities, respectively.
From Fig.~\ref{Figure:TotalUtility_correlated} and Fig.~\ref{Figure:TotalUtility_uncorrelated}, we also observe that Prop-LowComp-$t$ significantly outperforms all the baseline schemes in case-$t$ for $t=$ pp and ip;
Prop-up outperforms all the baseline schemes in case-up.
The results for large $K$ and $N$ shown in Fig.~\ref{Figure:TotalUtility_correlated} and Fig.~\ref{Figure:TotalUtility_uncorrelated} are similar to those for small $K$ and $N$ shown in Fig.~\ref{Figure_Throughput_small_K_correlated} and Fig.~\ref{Figure_Throughput_small_K_uncorrelated}.


\section{Conclusion}
This paper considered the joint optimization of preamble selection and access barring for random access in MTC under an arbitrary joint device activity distribution that can reflect dependent and independent device activities.
We considered the cases of perfect, imperfect, and unknown general joint device activity distributions and formulated the average, worst-case average, and sample average throughput maximization problems, respectively. All three problems are challenging nonconvex problems. 
We proposed iterative algorithms with convergence guarantees to tackle these problems with various optimization techniques.
Numerical results showed that the proposed solutions achieve significant gains over existing schemes in all three cases for both dependent and independent device activities, and the proposed low-complexity algorithms for the first two cases already achieve competitive performance.



\vspace{-1mm}

\section*{Appendix A: Proof for (\ref{T_Avg.New}) }
Substituting (\ref{T_x}) into (\ref{T_def}), we have~\cite{LWarxiv}\footnote{We omit some details due to
the page limitation. The complete proofs can be found in~\cite{LWarxiv}.}
\begin{align}
	\begin{array}{ll}
& 	\bar T\left( {\bf A} ,\epsilon ,\bf p \right)   = 	\sum\limits_{{\bf x \in \mathcal X}}
	p_{\bf x}  \sum\limits_{n\in \mathcal N}
	\sum\limits_{k\in \mathcal K} x_k a_{k,n} \epsilon \prod\limits_{m\in {\mathcal K:m \neq k}} (1-x_m a_{m,n}\epsilon )    \quad \quad\quad \quad \quad \quad\quad \quad   \quad \quad\quad  \quad \  \quad\nonumber \\
	 \end{array}
\end{align}
\begin{align}
\begin{array}{ll}
	& = \sum\limits_{{\bf x \in \mathcal X}}
	p_{\bf x} \sum\limits_{n\in \mathcal N}
	\sum\limits_{ k \in \mathcal K }  {x}_{k} a_{k,n}  \epsilon  \sum\limits_{
		\mathcal K' \subseteq {\mathcal K}\setminus \{k\}}
	 \prod\limits_{m \in \mathcal K'}(- {x}_{m} a_{m,n} \epsilon)  \nonumber \\	
	& =   -\sum\limits_{{\bf x \in \mathcal X}}
	p_{\bf x} \sum\limits_{n\in \mathcal N}
	\sum\limits_{ \mathcal K_1 \cup \mathcal K_2 \subseteq \mathcal K:  \mathcal K_1 \bigcap \mathcal K_2 =\emptyset, |\mathcal K_1|=1}  \prod\limits_{m \in \mathcal K_1}(- {x}_{m} a_{m,n} \epsilon) \prod\limits_{\ell \in \mathcal K_2}(- {x}_{\ell} a_{\ell,n} \epsilon) \\
  &	 =  - \sum\limits_{{\bf x \in \mathcal X}}
	p_{\bf x} \sum\limits_{n\in \mathcal N}
	\sum\limits_{
		\mathcal K' \subseteq {\mathcal K}}
	|\mathcal K'|\prod\limits_{k \in \mathcal K'}(- {x}_{k} a_{k,n} \epsilon)   
=	m(-1)^{m-1} \epsilon^m\sum\limits_{n\in \mathcal N}\sum\limits_{
		\mathcal K' \subseteq {\mathcal K}:|\mathcal K' |=m}  \left(\sum\limits_{{\bf x \in \mathcal X}}
	p_{\bf x}
	\prod\limits_{k \in \mathcal K'}{x}_{k}  \right)	\prod\limits_{k \in \mathcal K'}a_{k,n}.  \nonumber
	\end{array}
\end{align}

\vspace{-4mm}
\section*{Appendix B: Proof of Theorem \ref{Thm_OptimmalSolutionBCD} }

First, it is clear that each problem in (\ref{Prob:Perfectcase_RAO}) has the same form as the problem in
\cite[Exercise 4.8]{Boyd2004convex}.
According to the analytical solution of the problem in \cite[Exercise 4.8]{Boyd2004convex}, a set of optimal points of each problem in~(\ref{Prob:Perfectcase_RAO}) are given by~(\ref{PerUpdate_a}).
Next, since $ \bar T ({\bf A}, \epsilon,{\bf p})$ is a polynomial with respect to $\epsilon$, by checking all roots of $\frac{\partial \bar T ({\bf A}, \epsilon,{\bf p})}{\partial \epsilon}=0$ and the endpoints of the interval, we can obtain the set of optimal points of the problem in~(\ref{Prob:Perfectcase_ACB}), which is given by~(\ref{PerUpdate_epsilon}).
Therefore, we complete the proof of Theorem~\ref{Thm_OptimmalSolutionBCD}.

\section*{Appendix C: Complexity analysis for applying Theorem \ref{Thm_OptimmalSolutionBCD}}
 As constants $  m  ^2(-1)^{m-1} $, $m(-1)^{m-1}$, $m\in \mathcal K$ and $\sum\nolimits_{{\bf x \in \mathcal X}}
p_{\bf x}
\prod\nolimits_{k \in \mathcal K'}{x}_{k}$, ${\mathcal  K'} \subseteq \mathcal K$ are computed in advance,
the corresponding computational complexities are not considered below.
First, we analyze the computational complexity for determining the set in~(\ref{PerUpdate_a}).
In each iteration, we first compute $m(-1)^{m-1}\epsilon^{m}$, $m\in \mathcal K$ in $2K $ flops.
Then, we compute $Q_{k,n}({\bf a}_{-k}, \epsilon, {\bf p}  ) $, $k \in \mathcal K$, $n \in \mathcal N$.
In~(\ref{Q_kn}), the summation with respect to $m$ contains $K$ summands, the summation with respect to $\mathcal K' $ contains $\binom{K-1}{m-1}$ summands, and $   \prod\nolimits_{\ell\in \mathcal K' : \ell \neq k  }  a_{\ell,n}$ for $\mathcal K' \subseteq \mathcal K$ with $k \in \mathcal K',| \mathcal K'| =m $ involves $m-2$ multiplications.
Thus, calculating $Q_{k,n}({\bf a}_{-k}, \epsilon, {\bf p}  ) $, $k \in \mathcal K$, $n \in \mathcal N$ costs
$
\left(K-1 + \sum\nolimits_{m \in \mathcal K} \left((m -2+1)\binom{K-1}{m-1}+1 + \binom{K-1}{m-1}-1 \right) \right)NK + 2K	=NK(K+1)(2^{(K-2) }+1)-2K(N-1)
$
flops, i.e., has computational complexity $   \mathcal O \left(  N K^2 2^K\right)$.
For all $k \in \mathcal K$, the computational complexity for finding the largest one among $ Q_{k,n}({\bf a}_{-k}, \epsilon, {\bf p}  ), n\in \mathcal N$ is $\mathcal O (N)$.
Thus, the overall computational complexity for determining the sets in (\ref{PerUpdate_a}) is $ \mathcal O \left(     N K^2 2^K   \right) + \mathcal O \left(     KN  \right) = \mathcal O \left(  N K^2 2^K\right)$.

Then, we analyze the computational complexity for determining the set in (\ref{PerUpdate_epsilon}).
Note that $	q({\bf A},\epsilon,{\bf p})  $ in~(\ref{q}) is a univariate polynomial of order $K-1$ with respect to $\epsilon$ for any given $\bf A$ and $\bf p$.
We first calculate the coefficients of $	q({\bf A},\epsilon,{\bf p}) $.
In~(\ref{q}), the summation with respect to $m$ contains $K$ summands, the summation with respect to $n$ and $\mathcal K'   $ contains $N\binom{K }{m }$ summands, and $   \prod\nolimits_{\ell\in \mathcal K'    }  a_{\ell,n}$ for $n \in \mathcal N$ and $\mathcal K' \subseteq \mathcal K$ with $| \mathcal K'| =m $ involves $m-1$ multiplications.
Thus, calculating the coefficients of $	q({\bf A},\epsilon,{\bf p})   $ costs
$
\sum\nolimits_{m \in \mathcal K} \left((m -1+1) N\binom{K }{m }+1 + N\binom{K }{m }-1 \right) 	= N(K+2)2^{(K-1)}  -N
$
flops, i.e., has computational complexity $\mathcal O(NK2^K)$.
Next, we obtain the $K-1$ roots of $ 	q({\bf A},\epsilon,{\bf p}) $ based on its coefficients by using the QR algorithm with computational complexity $\mathcal O(K^3)$.
Furthermore, finding the real roots in $[0,1]$ from these $K-1$ roots has computational complexity $\mathcal O(K)$.
Hence, the computational complexity for determining $\mathcal B({\bf A},{\bf p}) $ is $\mathcal O (NK2^K  ) + \mathcal O (     K^3 )+ \mathcal O  (K)=  \mathcal O (NK2^K)$.
Analogously, we know that the computational complexity for computing $\bar T\left( {\bf A}, z,{\bf p}  \right)$, $z \in \mathcal B({\bf A},{\bf p}) \cup \{1\}  $  is $ \mathcal O \left(  NK^2 2^K   \right) $.
The computational complexity for finding the largest ones among $\bar T\left( {\bf A}, z,{\bf p}  \right)$, $z \in \mathcal B({\bf A},{\bf p}) \cup \{1\}  $ is $\mathcal O (K)$.	
Therefore, the overall computational complexity for determining the set in (\ref{PerUpdate_epsilon}) is $\mathcal O(   NK2^K   )+\mathcal O(    NK^22^K   )+\mathcal O(     K  ) =\mathcal O(  NK^22^K  ) $.

\section*{Appendix D: Proof of Theorem \ref{Thm_BCD} }
First, we show that Algorithm~\ref{Alg_PBCD} stops in a finite number of iterations.
Let $\left({\bf A}^{(i)},\epsilon^{(i)}\right)$ denote the preamble selection distributions and the access barring factor obtained at iteration $i$.
 $\mathcal A  \triangleq \left\{ {\bf A} :  {\bf a}_{k}= {\bf e}_{n_k},  n_k\in \mathcal N, k\in\mathcal K \right\}$ contains $N^K$ elements, $\mathcal B({\bf A},{\bf p}) \cup \{1\} $ contains no more than $K$ elements.
Thus,
$
	\mathcal C  \triangleq \left\{ ({\bf A}, \epsilon  ) :  {\bf A} \in\mathcal A,  \epsilon \in \mathcal B({\bf A},{\bf p})\cup\{1\} \right\}
$
and
$
\mathcal T  \triangleq \left\{  \bar T \left( {\bf A}, \epsilon, {\bf p} \right):    \left( {\bf A}, \epsilon\right) \in \mathcal C   \right\}
$
both contain no more than $KN^K$ elements.
In addition, as $\bar T \left({\bf A}^{(i)},\epsilon^{(i)},{\bf p}\right)$ is nondecreasing with $i$ (due to the block coordinate optimizations in each iteration) and $\bar T \left({\bf A}^{(i)},\epsilon^{(i)},{\bf p}\right) \in \mathcal T$, $i\geq1$, by contradiction, we can show that there exists an integer $d \leq KN^K$ such that
$  \bar T\left({\bf A }^{(d)}, \epsilon^{(d)},{\bf p } \right)= \bar T\left({\bf A }^{(d+1)},   \epsilon^{(d+1)},{\bf p } \right)$.
According to Steps $4$ - $11$ in Algorithm~\ref{Alg_PBCD}, we have $\left({\bf A }^{(d+1)},   \epsilon^{(d+1)}  \right)  =  \left({\bf A }^{(d)},   \epsilon^{(d)}   \right)$, which satisfies the stopping criteria of Algorithm~\ref{Alg_PBCD}.
Thus, Algorithm~\ref{Alg_PBCD} stops at iteration $d+1$, which is no more than $NK^N+1$ iterations.
Next, we show that Algorithm~\ref{Alg_PBCD} returns a stationary point of Problem~\ref{Prob:Perfectcase}.
By $  \left ({\bf A }^{(d+1)},   \epsilon^{(d+1)}  \right) =  \left({\bf A }^{(d)},   \epsilon^{(d)}   \right)$ and Theorem~\ref{Thm_OptimmalSolutionBCD}, ${\bf a}_{k}^{(d)}$ is the optimal point of the problem in~(\ref{Prob:Perfectcase_RAO}) with
${\bf a}_{-k}=  {\bf a}_{-k} ^{(d)}$ and $\epsilon = \epsilon^{(d)}$, for $k \in \mathcal K$, and $\epsilon^{(d)}$ is the optimal point of the problem in~(\ref{Prob:Perfectcase_ACB}) with ${\bf A}=  {\bf A}^{(d)}$.
In addition, note that the objective function $\bar T({\bf A},\epsilon,{\bf p})$ is continuously differentiable and the feasible set of Problem~\ref{Prob:Perfectcase} is convex.
Thus, according to the proof for~\cite[Proposition 2.7.1]{Bertsekas1998NP},
we know that $\left({\bf A }^{(d)},   \epsilon^{(d)}   \right)$ satisfies the first-order optimality condition of Problem~\ref{Prob:Perfectcase}, i.e., it is a stationary point.
Therefore, we complete the proof of Theorem~\ref{Thm_BCD}.

\section*{Appendix E: Proof of Theorem \ref{Lem_Prop_Opt}}
Let ${\bf f}\!\!:\mathbb{R}^{KN+1}\rightarrow \mathbb{R}^{KN+1}$ denote the vector mapping corresponding to one iteration of Algorithm~\ref{Alg_PBCD}.
	Let $ \left({\bf A}^{\diamond},\epsilon^{\diamond}\right)$ denote an optimal point of Problem~\ref{Prob:Perfectcase}.
	We construct a feasible point $ \left({\bf A}^{\star},\epsilon^{\star}\right) \triangleq {\bf f} \left({\bf A}^{\diamond},\epsilon^{\diamond}\right)$ which satisfies the optimality property in Theorem~\ref{Lem_Prop_Opt}.
In the following, we show that $\left({\bf A}^{\star},\epsilon^{\star}\right)$ is also an optimal point of Problem~\ref{Prob:Perfectcase}.
	According to Theorem~\ref{Thm_OptimmalSolutionBCD}, we have $ \bar T \left( {\bf A}^{\diamond}, \epsilon^{\diamond}, {\bf p}\right) \leq \bar T \left( {\bf a}_1^{\star}, {\bf a}_2^{\diamond}, ..., {\bf a}_K^{\diamond},     \epsilon^{\diamond}, {\bf p}\right)
	\leq
	\bar T \left( {\bf a}_1^{\star}, {\bf a}_2^{\star}, ..., {\bf a}_K^{\diamond},     \epsilon^{\diamond}, {\bf p}\right)
	\leq ...
	\leq
		\bar T \left( {\bf a}_1^{\star}, {\bf a}_2^{\star}, ..., {\bf a}_K^{\star},     \epsilon^{\diamond}, {\bf p}\right)
	\leq	
	\bar T \left( {\bf A}^{\star},    \epsilon^{\star}, {\bf p}\right)	 $.
In addition, as $ \left( {\bf A}^{\diamond},\epsilon^{\diamond}\right)$ is an optimal point, and $ \left( {\bf A}^{\star},\epsilon^{\star}\right)$ is a feasible point, we have $ \bar T \left( {\bf A}^{\star},\epsilon^{\star}, {\bf p}\right) \leq  \bar T \left( {\bf A}^{\diamond}, \epsilon^{\diamond}, {\bf p}\right) $.
By $ \bar T \left( {\bf A}^{\star},\epsilon^{\star}, {\bf p}\right) \geq  \bar T \left( {\bf A}^{\diamond}, \epsilon^{\diamond}, {\bf p}\right)$ and $\bar T \left( {\bf A}^{\star},\epsilon^{\star}, {\bf p}\right) \leq  \bar T \left( {\bf A}^{\diamond}, \epsilon^{\diamond}, {\bf p}\right)$, we have $\bar T \left( {\bf A}^{\star},\epsilon^{\star}, {\bf p}\right) =  \bar T \left( {\bf A}^{\diamond}, \epsilon^{\diamond}, {\bf p}\right)$, which implies that $\left( {\bf A}^{\star},\epsilon^{\star}\right)$ is also optimal.
	Therefore, we complete the proof of Theorem~\ref{Lem_Prop_Opt}.

\vspace{-2mm}
\section*{Appendix F: Proof of Lemma \ref{Lem_ImpefectHL_Equi} }
The inner problem of Problem~\ref{Prob:Imperfectcase}, $\min_{{\bf p} \in \mathcal P} \bar T({\bf A},\epsilon,{\bf p} )$, is a feasible LP with respect to $\bf p$ for all $({\bf A}, \epsilon)$ satisfying (\ref{C1}), (\ref{C12}),
(\ref{C2}) and (\ref{C3}).
As strong duality holds for LP, the primal problem, $\min_{{\bf p} \in \mathcal P} \bar T({\bf A},\epsilon,{\bf p} )$, and its dual problem, share the same optimal value.
Furthermore, by eliminating the equality constraints in the dual problem, we can equivalently convert the dual problem to the following problem\cite{LWarxiv}
 \par\vspace{-20pt}{{\small \begin{align}
\max\limits_{{\boldsymbol{\lambda}}\succeq0,\nu  }& \quad
 \Big(  \sum\limits_{{\bf x}\in \mathcal X  }\underline{p}_{\bf x}  - 1\Big)\nu+
  \sum\limits_{{\bf x}\in \mathcal X  } \Big( (\underline{p}_{\bf x} - \overline{p}_{\bf x})\lambda_{\bf x}
 +\underline{p}_{\bf x}
  \sum\limits_{n\in \mathcal N}
     \sum\limits_{k\in \mathcal K} x_k a_{k,n}\epsilon \prod\limits_{\ell\in {\mathcal K}: \ell\neq k} (1- x_{\ell} a_{\ell,n}  \epsilon )     \Big)
 \label{Prob:EQPofInnerDual} \\
\mathrm{s.t.} & \quad \nu +\lambda_{\bf x}+ \sum\limits_{n\in \mathcal N}
     \sum\limits_{k\in \mathcal K} x_k a_{k,n}\epsilon \prod\limits_{\ell\in {\mathcal K}: \ell\neq k} (1-x_{\ell}a_{\ell,n}\epsilon ) \geq 0 ,
     \ {\bf x} \in \mathcal X. \nonumber
 \end{align}}%
Thus, the problems in (\ref{Prob:EQPofInnerDual}) and $\min_{{\bf p} \in \mathcal P} \bar T({\bf A},\epsilon,{\bf p} )$ share the same optimal value,
which can be viewed as a function of $({\bf A}, \epsilon)$.
Thus, the
max-min problem in Problem \ref{Prob:Imperfectcase} can be equivalently converted to the maximization problem in
Problem \ref{Prob:ImperfectcaseEQP}, by replacing the inner problem, $\min_{{\bf p} \in \mathcal P} \bar T({\bf A},\epsilon,{\bf p} )$, with the problem in~(\ref{Prob:EQPofInnerDual}), and replacing $\epsilon a_{k,n}$ with new variables $b_{k,n}$ in both the objective function and constraint functions.
Therefore, we complete the proof of Lemma~\ref{Lem_ImpefectHL_Equi}.

\vspace{-3mm}
\section*{Appendix G: Proof of Theorem \ref{Thm_Ropt} }
We show that the assumptions in \cite[Theorem 1]{razaviyayn2014successive} are satisfied.
 It is clear that
 $\tilde h({\bf B},{\boldsymbol{\lambda}}, {\nu}, {\bf B}^{(s)}) $
  and
  $\tilde h_{\text{c}} ({\bf B},{\boldsymbol{\lambda}},\epsilon  ,{\nu},{\bf B}^{(s)},{\bf x} ) $, ${\bf x} \in \mathcal X$
  are continuously differentiable with respect to $({\bf B},{\boldsymbol{\lambda}},\epsilon,{\nu} )$,
  and each of them is the sum of linear functions and strongly concave functions.
Thus, $\tilde h({\bf B},{\boldsymbol{\lambda}}, {\nu}, {\bf B}^{(s)}) $
 	and
 	$\tilde h_{\text{c}} ({\bf B},{\boldsymbol{\lambda}},\epsilon  ,{\nu},{\bf B}^{(s)},{\bf x} ) $, ${\bf x} \in \mathcal X$ satisfy the first, second, and third assumptions.
From (\ref{Approximation_Imperfectcase_h}) and (\ref{Approximation_Imperfectcase_hx}), it is clear that $\tilde h({\bf B},{\boldsymbol{\lambda}}, {\nu}, {\bf B}^{(s)}) $
	and
	$\tilde h_{\text{c}} ({\bf B},{\boldsymbol{\lambda}},\epsilon  ,{\nu},{\bf B}^{(s)},{\bf x} ) $, ${\bf x} \in \mathcal X$ satisfy the fourth and fifth assumptions.}
 Let $\nabla_{\bf B}^2  T({{\bf B}},1, {\bf x})
\triangleq
\big(\frac{\partial^2  T({{\bf B}}, 1, {\bf x}) }{\partial b_{k,n}b_{k',n'}}\big)_{(k,n),(k',n') \in \mathcal K \times \mathcal N }  $
and
$
	\nabla_{\bf B}^2 \sum\nolimits_{{\bf x} \in \mathcal X} \underline{p}_{\bf x}T({{\bf B}},1, {\bf x})
\triangleq
\big(  \sum\nolimits_{{\bf x} \in \mathcal X} \underline{p}_{\bf x} \frac{   \partial^2 T({{\bf B}}, 1, {\bf x}) }{\partial b_{k,n}b_{k',n'}}\big)_{(k,n),(k',n') \in \mathcal K \times \mathcal N }
$
 {denote} the Hessians of $T({{\bf B}},1, {\bf x}) $ and $\sum\nolimits_{{\bf x} \in \mathcal X} \underline{p}_{\bf x}T({{\bf B}},1, {\bf x})$, respectively.
We can show $\nabla_{\mathbf B}^2  T({{\bf B}},1, {\bf x}) \preceq  $ $ \sqrt{N    }     \|\mathbf x\|_1^2  {\bf E}$, ${\bf x}\in\mathcal X$,
and
$
 	\nabla_{\bf B}^2 \sum\nolimits_{{\bf x} \in \mathcal X} \underline{p}_{\bf x}T({{\bf B}},1, {\bf x})  \preceq \sqrt{N  \sum\nolimits_{k \in \mathcal K} \sum\nolimits_{k' \in \mathcal K: k'\neq k } \big(\sum\nolimits_{\mathbf x \in \mathcal X} \underline p_{\mathbf x} x_k x_{k'}  \|\mathbf x\|_1  \big)^2}   {\bf E} 
$~\cite{LWarxiv}, where $\bf E$ represents the identity matrix.
 Thus, using the inequality implied by Taylor's theorem~\cite[Eq. (25)]{SunMM},
 we know that $\tilde h({\bf B},{\boldsymbol{\lambda}}, {\nu}, {\bf B}^{(s)}) $
 and
 $\tilde h_{\text{c}} ({\bf B},{\boldsymbol{\lambda}},\epsilon  ,{\nu},{\bf B}^{(s)},{\bf x} ) $, ${\bf x} \in \mathcal X$ satisfy the sixth assumption.
Therefore, Theorem~\ref{Alg_SPSCA} readily follows from~\cite[Theorem 1]{razaviyayn2014successive}.

\vspace{-4mm}
\section*{Appendix H: Proof of Lemma \ref{Thm_SubOptLB} }
For all $({\bf A},\epsilon)$ satisfying (\ref{C1}), (\ref{C12}), (\ref{C2}), (\ref{C3}), as ${\bf p} \in \tilde {\mathcal P} $,
	we have $
		\min\nolimits_{ {\bf p} \in\tilde {\mathcal P}} \tilde T ( {\bf A},{\epsilon}, {\bf p} ) \geq
		\epsilon  \sum\nolimits_{k\in \mathcal {K}} \sum\nolimits_{{\bf x} \in \mathcal X }
		{\underline p}_{\bf x}  x_k
		-\epsilon^2 \sum\nolimits_{n\in \mathcal {N}} \sum\nolimits_{k\in \mathcal {K }}a_{k,n} \sum\nolimits_{\ell \in {\mathcal K}: \ell > k}a_{\ell,n}
		\sum\nolimits_{{\bf x} \in \mathcal X }{\overline p}_{\bf x} x_kx_{\ell}
	 $,
where the equality holds when
	\begin{align}
		\begin{array}{ll}
   \sum\limits_{{\bf x} \in \mathcal X} p_{\bf x} =1, \ 
	  \sum\limits_{{\bf x} \in \mathcal X }
		{ p}_{\bf x}x_k
		=\sum\limits_{{\bf x} \in \mathcal X }\underline{ p}_{\bf x}x_k,
		\ k \in \mathcal K,  \ 
		 \sum\limits_{{\bf x} \in \mathcal X }{ p}_{\bf x} x_kx_{\ell}
		=
		\sum\limits_{{\bf x} \in \mathcal X }\overline{ p}_{\bf x} x_kx_{\ell} , \ k,\ell \in \mathcal K, k< \ell  \end{array} \label{AppendGEquality}
	\end{align}%
are satisfied.
Thus, to show Lemma~\ref{Thm_SubOptLB}, it is equivalent to show that the system of linear equations with variable $\bf p$ in~(\ref{AppendGEquality}) has a solution.
Let $g({\bf x}) \triangleq 1+  \sum\nolimits_{k \in \mathcal K} x_k 2^{k-1} $~\cite{Teugels1990}.
Obviously, $g: \mathcal X \rightarrow  \{1,2,3,...,2^K\}$ is a bijection.
Thus, for all ${\bf x} \in \mathcal X$, we can also write $p_{\bf x}$ as $p_{g({\bf x})}$.
As $g(\bf 0)$, $\min \{ g({\bf x}) |    x_k =1, {\bf x} \in \mathcal X \} =g({\bf e}_k) $, $k \in \mathcal K$, and $  \min \{ g({\bf x}) |    x_k =1,x_{\ell}= 1, {\bf x} \in \mathcal X \} =g({\bf e}_k + {\bf e}_{\ell}  )$, $k,\ell \in \mathcal K, k<\ell $ are all different, the matrix of the system of linear equations in~(\ref{AppendGEquality}) has full row rank.
In addition, note that the number of equations, $ \frac{K^2+ K+ 2}{2} $, is no greater than the number of variables, $2^K$.
Thus, the system of linear equations in~(\ref{AppendGEquality}) has a solution.
 We complete the proof of Lemma~\ref{Thm_SubOptLB}.

\vspace{-5mm}
 
\section*{Appendix I: Proof of Theorem \ref{Thm_SPSCA} }
	
For all $k\in\mathcal K$ and all $t \geq 1$, let $n^{(t)} \triangleq \mathop{\arg\max}_{n \in \mathcal N} c_{k,n}^{(t)} $, which is a singleton, and let $n^{'(t)}$ denote one index in $\mathop{\arg\max}_{n \in \mathcal N: n \neq n^{(t)}} c_{k,n}^{(t)}$.	
First, we show that for any $0<\omega < \frac{1}{4}\big(c_{k, n^{(t)}}^{(t)}-c_{k, n^{'(t)}}^{(t)}\big) $, the optimization problem in~(\ref{Prob:Unknowncase_RAO}) and the following QP (which is strongly convex),
\par\vspace{-15pt}{{\small	\begin{align}
			\max\limits_{{\bf a}_k  }&  \quad   Q_{\omega}^{(t)}\left({\bf a}_k\right) \triangleq  \tilde  T_{\text{st},k}^{(t)} \big({\bf a}_k,{\bf A}^{(t-1)}  , \epsilon^{(t-1)}  \big )
- \omega \sum\nolimits_{n\in\mathcal N} \big( a_{k,n} - a_{k,n}^{(t-1)}  \big)^2  , \quad   k \in \mathcal K    \label{EQQP} \\
			\mathrm{s.t.}&  \quad  \text{(\ref{C2})}, \ \text{(\ref{C3})},\nonumber
	\end{align}}%
share the same optimal point.
According to Theorem~\ref{Thm_OptimmalSolutionSPSCA}, the optimal point of the problem in~(\ref{Prob:Unknowncase_RAO}) is ${\bf e}_{n^{(t)}}$.
In addition, for any ${\bf a}_k$ satisfying the constraints in (\ref{C2}) and (\ref{C3}), we can show
$
		     Q_{\omega}^{(t)}  \left({\bf a}_k \right)   \leq  Q_{\omega}^{(t)} \left({\bf e}_{n^{(t)}} \right)  
		     $~\cite{LWarxiv}.
By noting that the QP in~(\ref{EQQP}) is strongly convex (as $\omega >0$), ${\bf e}_{n^{(t)}}$ is also the unique optimal point of the strongly convex QP.
Thus, if $\mathop{\arg\max}_{n\in\mathcal N}c_{k,n}^{(t)}$ is a singleton for all $k\in\mathcal K$ and all $t \geq 1$, Algorithm~\ref{Alg_SPSCA} can be viewed as stochastic parallel SCA in \cite{Yang2016parallel}, for solving the following stochastic problem
\par\vspace{-15pt}{{\small\begin{align}
\max\limits_{{\bf A},\epsilon}& \quad  \mathbb E  \left[	\frac{M}{I} \sum\limits_{m \in \mathcal M} \text{I}[  m= \xi ]  \sum\limits_{i \in \mathcal I_m }T \left({\bf A},{\epsilon},{\bf x}_i \right ) \right]    \label{EQPUnknonw} \\
\mathrm{s.t.}& \quad  \text{(\ref{C1})},\ \text{(\ref{C12})},  \ \text{(\ref{C2})}, \  \text{(\ref{C3})},   \nonumber
\end{align}}%
 where $\xi $ (index of mini-batch) follows the uniform distribution over $\mathcal M$, and the expectation is taken over $\xi$.
 By noting that $\mathbb E  \left[	\frac{M}{I} \sum\nolimits_{m \in \mathcal M} \text{I}[  m= \xi ]  \sum\nolimits_{i \in \mathcal I_m } T \left({\bf A},{\epsilon},{\bf x}_i \right ) \right] = \bar T_{\text{st}}\left( {\bf A}, \epsilon \right)$, the stochastic problem in~(\ref{EQPUnknonw}) is equivalent to Problem~\ref{Prob:Unknowncase}.
Next, we show the convergence of Algorithm~\ref{Alg_SPSCA} by showing that the assumptions in \cite[Theorem 1]{Yang2016parallel} are satisfied.
 Obviously, the constraint set of Problem~\ref{Prob:Unknowncase} is compact and convex.
It is clear that for any $\xi \in \mathcal M$, $ \frac{M}{I} \sum\nolimits_{m \in \mathcal M} \text{I}[  m= \xi ]  \sum\nolimits_{i \in \mathcal I_m}T \left({\bf A},{\epsilon},{\bf x}_i \right )$ is smooth on the constraint set of Problem \ref{Prob:Unknowncase},
and hence it is continuously differentiable and its derivative is Lipschitz continuous. 
Random variables ${\xi}^{(0)}, {\xi}^{(1)},...$ are bounded and identically distributed. 
Therefore, Theorem \ref{Thm_SPSCA} readily follows from\text{\cite[Theorem 1]{Yang2016parallel}.}





\IEEEpeerreviewmaketitle

\ifCLASSOPTIONcaptionsoff
  \newpage
\fi



\end{document}